\documentclass[zpreprint,zbstdefault]{zeus_paper}
\usepackage[english]{babel}

\newcommand{\ZcoosysB}{%
The ZEUS coordinate system is a right-handed Cartesian system, with the $Z$
axis pointing in the proton beam direction, referred to as the ``forward
direction'', and the $X$ axis pointing towards the centre of HERA.
The coordinate origin is at the nominal interaction point.\xspace}
\newcommand{\Zpsrap}{%
The pseudorapidity is defined as $\eta=-\ln\left(\tan\frac{\theta}{2}\right)$,
where the polar angle, $\theta$, is measured with respect to the proton beam
direction. The azimuthal angle in the $X$-$Y$ plane is called $\phi$.\xspace}

\newcommand{\ZcoosysfnBeta}{\footnote{\ZcoosysB\Zpsrap}}
\newcommand{\Zdetdesc}{%
A detailed description of the ZEUS detector can be found 
elsewhere~\cite{zeus:1993:bluebook}. A brief outline of the 
components most relevant for this analysis is given
below.\xspace}
\newcommand{\Zctddesc}[1]{%
Charged particles were tracked in the central tracking detector (CTD)~\citeCTD,
which operated in a magnetic field of $1.43\Tesla$ provided by a thin 
superconducting coil. The CTD consisted of 72~cylindrical drift chamber 
layers, organised in 9~superlayers covering the polar-angle#1 region 
\mbox{$15^\circ<\theta<164^\circ$}. }

\newcommand{\Zcaldesc}{%
The high-resolution uranium--scintillator calorimeter (CAL)~\citeCAL consisted 
of three parts: the forward (FCAL), the barrel (BCAL) and the rear (RCAL)
calorimeters. Each part was subdivided transversely into towers and
longitudinally into one electromagnetic section (EMC) and either one (in RCAL)
or two (in BCAL and FCAL) hadronic sections (HAC). The smallest subdivision of
the calorimeter is called a cell.  The CAL energy resolutions, as measured under
test-beam conditions, are $\sigma(E)/E=0.18/\sqrt{E}$ for electrons and
$\sigma(E)/E=0.35/\sqrt{E}$ for hadrons, with $E$ in $\Gev$.}

\chardef\usc=95
\chardef\til=126
\catcode`\@=11 
\DeclareRobustCommand\xdotspace{\futurelet\@let@token\@xdotspace}
\def\@xdotspace{%
  \ifx\@let@token.\else
  \ifx\@let@token\bgroup.\else
  \ifx\@let@token\egroup.\else
  \ifx\@let@token\/.\else
  \ifx\@let@token\ .\else
  \ifx\@let@token~.\else
  \ifx\@let@token!.\else
  \ifx\@let@token,.\else
  \ifx\@let@token:.\else
  \ifx\@let@token;.\else
  \ifx\@let@token?.\else
  \ifx\@let@token/.\else
  \ifx\@let@token'.\else
  \ifx\@let@token).\else
  \ifx\@let@token-.\else
  \ifx\@let@token\@xobeysp.\else
  \ifx\@let@token\space.\else
  \ifx\@let@token\@sptoken.\else
   .\space
   \fi\fi\fi\fi\fi\fi\fi\fi\fi\fi\fi\fi\fi\fi\fi\fi\fi\fi}
\catcode`\@=12 

\newcommand{\stru}[2]{%
   \relax\ifmmode\hbox{\vrule height#1 depth#2 width0pt}%
   \else\vrule height#1 depth#2 width0pt\fi}

\newcommand{\Ronum}[1]{\uppercase\expandafter{\romannumeral#1}}
\newcommand{\ronum}[1]{\expandafter{\romannumeral#1}}
\DeclareRobustCommand{\LaTeXZ}{%
  \LaTeX\kern-.05em4\kern-.1em
  {\raisebox{-0.2ex}{$\scriptstyle\text{ZEUS}$}}\xspace}

\DeclareMathAlphabet{\mathbf}{OT1}{cmr}{bx}{sl}
\newcommand{\eVdist}{\kern-0.06667em}

\newcommand{\Gev}{{\text{Ge}\eVdist\text{V\/}}}

\newcommand{\mev}{{\,\text{Me}\eVdist\text{V\/}}}
\newcommand{\gev}{{\,\text{Ge}\eVdist\text{V\/}}}

\newcommand{\pbi}{\,\text{pb}^{-1}}

\newcommand{\met}{\,\text{m}}

\newcommand{\cm}{\,\text{cm}}

\newcommand{\Tesla}{\,\text{T}}

\newcommand{\slashfrac}[2]{%
  \raisebox{0.5ex}{\ensuremath #1}\kern-0.12em/\kern-0.08em
  \raisebox{-.8ex}{\ensuremath #2}}

\newcommand{\sqr}[3]{%
    {\vcenter{\hrule height.#3ex\hbox{\vrule width.#2ex height#1ex
     \kern#1ex\vrule width.#3ex}\hrule height.#2ex}}}

\catcode`\@=11 
\newcommand{\parenbar}{\mathpalette\p@renb@r}
\def\p@renb@r#1#2{\vbox{%
  \ifx#1\scriptscriptstyle \dimen@.7em\dimen@ii.2em\else
  \ifx#1\scriptstyle \dimen@.8em\dimen@ii.25em\else
  \dimen@1em\dimen@ii.4em\fi\fi \offinterlineskip
  \ialign{\hfill##\hfill\cr
    \vbox{\hrule width\dimen@ii}\cr
    \noalign{\vskip-.3ex}%
    \hbox to\dimen@{$\mathchar300\hfil\mathchar301$}\cr
    \noalign{\vskip-.3ex}%
    $#1#2$\cr}}}
\catcode`\@=12 

\newcommand{\IP}{{\rm I$\kern-0.01667em$P}\xspace}

\mathchardef\qsm=63
\mathchardef\pls=43
\mathchardef\mns=512
\mathchardef\plm=518
\mathchardef\eql=61
\mathchardef\smallleft=300
\mathchardef\smallright=301
\mathchardef\les=316
\mathchardef\gre=318
\mathchardef\leq=532
\mathchardef\grq=533

\catcode`\@=11 
\newcounter{pict@width}
\newcounter{pict@height}
\newlength{\pict@scale}
\newcommand{\psfigadd}[4]{%
\setcounter{pict@width}{1*\ratio{#2+\pict@scale/2}{\pict@scale}}
\setcounter{pict@height}{1*\ratio{#3+\pict@scale/2}{\pict@scale}}
\setlength{\unitlength}{\pict@scale}
\hbox to #2{\hspace{-\fill}\begin{picture}(\thepict@width,\thepict@height)
\put(0,0){\psfig{figure=#1,width=#2,height=#3,clip=}}
\SetScale{0.283466457}
\SetWidth{1.763889}
{#4}
\end{picture}}
}
\newcounter{pict@widthfst}
\newcounter{pict@widthscd}
\newcounter{pict@widthtot}
\newcommand{\psfigaddtwo}[7]{%
\setcounter{pict@widthfst}{1*\ratio{#2+\pict@scale/2}{\pict@scale}}
\setcounter{pict@widthscd}{1*\ratio{#2+#4+\pict@scale/2}{\pict@scale}}
\setcounter{pict@widthtot}{1*\ratio{#2+#4+#6+\pict@scale/2}{\pict@scale}}
\setcounter{pict@height}{1*\ratio{#3+\pict@scale/2}{\pict@scale}}
\setlength{\unitlength}{\pict@scale}
\hbox{\hspace{-\fill}\begin{picture}(\thepict@widthtot,\thepict@height)
\put(0,0){\psfig{figure=#1,width=#2,height=#3,clip=}}
\put(\thepict@widthscd,0){\psfig{figure=#5,width=#6,height=#3,clip=}}
\SetScale{0.283466457}
\SetWidth{1.763889}
{#7}
\end{picture}}
}
\newcommand{\psfigror}[4]{%
\setcounter{pict@width}{1*\ratio{#2+\pict@scale/2}{\pict@scale}}
\setcounter{pict@height}{1*\ratio{#3+\pict@scale/2}{\pict@scale}}
\setlength{\unitlength}{\pict@scale}
\hbox{\begin{picture}(\thepict@width,\thepict@height)
\put(0,\thepict@height){\psfig{figure=#1,width=#3,height=#2,clip=,angle=270}}
\SetScale{0.283466457}
\SetWidth{1.763889}
{#4}
\end{picture}}
}
\newcommand{\psfigrol}[4]{%
\setcounter{pict@width}{1*\ratio{#2+\pict@scale/2}{\pict@scale}}
\setcounter{pict@height}{1*\ratio{#3+\pict@scale/2}{\pict@scale}}
\setlength{\unitlength}{\pict@scale}
\hbox{\begin{picture}(\thepict@width,\thepict@height)
\put(0,0){\psfig{figure=#1,width=#3,height=#2,clip=,angle=90}}
\SetScale{0.283466457}
\SetWidth{1.763889}
{#4}
\end{picture}}
}
\catcode`\@=12 
\newlength\listtextwidth

\catcode`\@=11 
\newlength{\@tabfninsert}
\newlength{\@tabfnwidth}
\newcommand{\tabfootnote}[2]{%
  \setlength{\@tabfninsert}{0.8em}
  \setlength{\@tabfnwidth}{\textwidth}
  \addtolength{\@tabfnwidth}{-\@tabfninsert}
  \addtolength{\@tabfnwidth}{-0.4em}
  \noindent\makebox[\@tabfninsert][r]{\footnotesize$^{#1}$\hfil}\hfill%
  \parbox[t]{\@tabfnwidth}{\footnotesize #2\hfill}}
\catcode`\@=12 

\def\citeCTD{{\cite{%
nim:a279:290,*npps:b32:181,*nim:a338:254%
}}\xspace}

\def\citeCAL{{\cite{%
nim:a309:77,*nim:a309:101,*nim:a321:356,*nim:a336:23%
}}\xspace}

\begin{document}
\prepnum{DESY--10--064}

\title{
Measurement of $\mathbf{D^{+}}$ and $\mathbf{\Lambda_{c}^{+}}$ production\\
in deep inelastic scattering at HERA
}                                                       
                    
\author{ZEUS Collaboration}
\draftversion{post-reading}
\date{June 2010}

\abstract{
Charm production in deep inelastic scattering has been measured with the ZEUS detector at HERA using an integrated luminosity of $120\pbi$. The hadronic decay channels $D^{+}\to K^{0}_{S} \pi^{+}$, $\Lambda_{c}^{+}\to p K^{0}_{S}$ and $\Lambda_{c}^{+}\to \Lambda \pi^{+}$, and their charge conjugates, were reconstructed. The presence of a neutral strange hadron in the final state reduces the combinatorial background and extends the measured sensitivity into the low transverse momentum region. The kinematic range is $0 < p_{T}(D^{+}, \Lambda_{c}^{+}) < 10\gev$, $|\eta(D^{+},\Lambda_{c}^{+})| < 1.6$, $1.5 < Q^{2} < 1000\gev^{2}$ and $0.02 < y < 0.7$. Inclusive and differential cross sections for the production of $D^{+}$ mesons are compared to next-to-leading-order QCD predictions. The fraction of $c$ quarks hadronising into $\Lambda_{c}^{+}$ baryons is extracted.
}

\makezeustitle

%
%
%
%

\def\3{\ss}
\pagenumbering{Roman}
                                                   %
\begin{center}
{                      \Large  The ZEUS Collaboration              }
\end{center}

{\small


{\mbox H.~Abramowicz$^{44, ad}$, }
{\mbox I.~Abt$^{34}$, }
{\mbox L.~Adamczyk$^{13}$, }
{\mbox M.~Adamus$^{53}$, }
{\mbox R.~Aggarwal$^{7}$, }
{\mbox S.~Antonelli$^{4}$, }
{\mbox P.~Antonioli$^{3}$, }
{\mbox A.~Antonov$^{32}$, }
{\mbox M.~Arneodo$^{49}$, }
{\mbox V.~Aushev$^{26, z}$
% ...
}
\newpage


\makebox[3em]{$^{1}$}
\begin{minipage}[t]{14cm}
{\it Argonne National Laboratory, Argonne, Illinois 60439-4815, USA}~$^{A}$

\end{minipage}\\
\makebox[3em]{$^{2}$}
\begin{minipage}[t]{14cm}
{\it Andrews University, Berrien Springs, Michigan 49104-0380, USA}

\end{minipage}\\
\makebox[3em]{$^{3}$}
\begin{minipage}[t]{14cm}
{\it INFN Bologna, Bologna, Italy}~$^{B}$

\end{minipage}\\
\makebox[3em]{$^{4}$}
\begin{minipage}[t]{14cm}
{\it University and INFN Bologna, Bologna, Italy}~$^{B}$

\end{minipage}\\
\makebox[3em]{$^{5}$}
\begin{minipage}[t]{14cm}
{\it Physikalisches Institut der Universit\"at Bonn,
Bonn, Germany}~$^{C}$

\end{minipage}\\
\makebox[3em]{$^{6}$}
\begin{minipage}[t]{14cm}
{\it H.H.~Wills Physics Laboratory, University of Bristol,
Bristol, United Kingdom}~$^{D}$

\end{minipage}\\
\makebox[3em]{$^{7}$}
\begin{minipage}[t]{14cm}
{\it Panjab University, Department of Physics, Chandigarh, India}

\end{minipage}\\
\makebox[3em]{$^{8}$}
\begin{minipage}[t]{14cm}
{\it Calabria University,
Physics Department and INFN, Cosenza, Italy}~$^{B}$

\end{minipage}\\
\makebox[3em]{$^{9}$}
\begin{minipage}[t]{14cm}
{\it Institute for Universe and Elementary Particles, Chonnam National University,\\
Kwangju, South Korea}

\end{minipage}\\
\makebox[3em]{$^{10}$}
\begin{minipage}[t]{14cm}
{\it Jabatan Fizik, Universiti Malaya, 50603 Kuala Lumpur, Malaysia}~$^{E}$

\end{minipage}\\
\makebox[3em]{$^{11}$}
\begin{minipage}[t]{14cm}
{\it Nevis Laboratories, Columbia University, Irvington on Hudson,
New York 10027, USA}~$^{F}$

\end{minipage}\\
\makebox[3em]{$^{12}$}
\begin{minipage}[t]{14cm}
{\it The Henryk Niewodniczanski Institute of Nuclear Physics, Polish Academy of Sciences,\\
Cracow, Poland}~$^{G}$

\end{minipage}\\
\makebox[3em]{$^{13}$}
\begin{minipage}[t]{14cm}
{\it Faculty of Physics and Applied Computer Science, AGH-University of Science and \\
Technology, Cracow, Poland}~$^{H}$

\end{minipage}\\
\makebox[3em]{$^{14}$}
\begin{minipage}[t]{14cm}
{\it Department of Physics, Jagellonian University, Cracow, Poland}

\end{minipage}\\
\makebox[3em]{$^{15}$}
\begin{minipage}[t]{14cm}
{\it Deutsches Elektronen-Synchrotron DESY, Hamburg, Germany}

\end{minipage}\\
\makebox[3em]{$^{16}$}
\begin{minipage}[t]{14cm}
{\it Deutsches Elektronen-Synchrotron DESY, Zeuthen, Germany}

\end{minipage}\\
\makebox[3em]{$^{17}$}
\begin{minipage}[t]{14cm}
{\it INFN Florence, Florence, Italy}~$^{B}$

\end{minipage}\\
\makebox[3em]{$^{18}$}
\begin{minipage}[t]{14cm}
{\it University and INFN Florence, Florence, Italy}~$^{B}$

\end{minipage}\\
\makebox[3em]{$^{19}$}
\begin{minipage}[t]{14cm}
{\it Fakult\"at f\"ur Physik der Universit\"at Freiburg i.Br.,
Freiburg i.Br., Germany}~$^{C}$

\end{minipage}\\
\makebox[3em]{$^{20}$}
\begin{minipage}[t]{14cm}
{\it Department of Physics and Astronomy, University of Glasgow,
Glasgow, United Kingdom}~$^{D}$

\end{minipage}\\
\makebox[3em]{$^{21}$}
\begin{minipage}[t]{14cm}
{\it Department of Engineering in Management and Finance, Univ. of
the Aegean, Chios, Greece}

\end{minipage}\\
\makebox[3em]{$^{22}$}
\begin{minipage}[t]{14cm}
{\it Hamburg University, Institute of Exp. Physics, Hamburg,
Germany}~$^{C}$

\end{minipage}\\
\makebox[3em]{$^{23}$}
\begin{minipage}[t]{14cm}
{\it Imperial College London, High Energy Nuclear Physics Group,
London, United Kingdom}~$^{D}$

\end{minipage}\\
\makebox[3em]{$^{24}$}
\begin{minipage}[t]{14cm}
{\it Institute of Particle and Nuclear Studies, KEK,
Tsukuba, Japan}~$^{I}$

\end{minipage}\\
\makebox[3em]{$^{25}$}
\begin{minipage}[t]{14cm}
{\it Institute of Physics and Technology of Ministry of Education and
Science of Kazakhstan, Almaty, Kazakhstan}

\end{minipage}\\
\makebox[3em]{$^{26}$}
\begin{minipage}[t]{14cm}
{\it Institute for Nuclear Research, National Academy of Sciences, and
Kiev National University, Kiev, Ukraine}

\end{minipage}\\
\makebox[3em]{$^{27}$}
\begin{minipage}[t]{14cm}
{\it Kyungpook National University, Center for High Energy Physics, Daegu,
South Korea}~$^{J}$

\end{minipage}\\
\makebox[3em]{$^{28}$}
\begin{minipage}[t]{14cm}
{\it Institut de Physique Nucl\'{e}aire, Universit\'{e} Catholique de Louvain, Louvain-la-Neuve,\\
Belgium}~$^{K}$

\end{minipage}\\
\makebox[3em]{$^{29}$}
\begin{minipage}[t]{14cm}
{\it Departamento de F\'{\i}sica Te\'orica, Universidad Aut\'onoma
de Madrid, Madrid, Spain}~$^{L}$

\end{minipage}\\
\makebox[3em]{$^{30}$}
\begin{minipage}[t]{14cm}
{\it Department of Physics, McGill University,
Montr\'eal, Qu\'ebec, Canada H3A 2T8}~$^{M}$

\end{minipage}\\
\makebox[3em]{$^{31}$}
\begin{minipage}[t]{14cm}
{\it Meiji Gakuin University, Faculty of General Education,
Yokohama, Japan}~$^{I}$

\end{minipage}\\
\makebox[3em]{$^{32}$}
\begin{minipage}[t]{14cm}
{\it Moscow Engineering Physics Institute, Moscow, Russia}~$^{N}$

\end{minipage}\\
\makebox[3em]{$^{33}$}
\begin{minipage}[t]{14cm}
{\it Moscow State University, Institute of Nuclear Physics,
Moscow, Russia}~$^{O}$

\end{minipage}\\
\makebox[3em]{$^{34}$}
\begin{minipage}[t]{14cm}
{\it Max-Planck-Institut f\"ur Physik, M\"unchen, Germany}

\end{minipage}\\
\makebox[3em]{$^{35}$}
\begin{minipage}[t]{14cm}
{\it NIKHEF and University of Amsterdam, Amsterdam, Netherlands}~$^{P}$

\end{minipage}\\
\makebox[3em]{$^{36}$}
\begin{minipage}[t]{14cm}
{\it Physics Department, Ohio State University,
Columbus, Ohio 43210, USA}~$^{A}$

\end{minipage}\\
\makebox[3em]{$^{37}$}
\begin{minipage}[t]{14cm}
{\it Department of Physics, University of Oxford,
Oxford, United Kingdom}~$^{D}$

\end{minipage}\\
\makebox[3em]{$^{38}$}
\begin{minipage}[t]{14cm}
{\it INFN Padova, Padova, Italy}~$^{B}$

\end{minipage}\\
\makebox[3em]{$^{39}$}
\begin{minipage}[t]{14cm}
{\it Dipartimento di Fisica dell' Universit\`a and INFN,
Padova, Italy}~$^{B}$

\end{minipage}\\
\makebox[3em]{$^{40}$}
\begin{minipage}[t]{14cm}
{\it Department of Physics, Pennsylvania State University, University Park,\\
Pennsylvania 16802, USA}~$^{F}$

\end{minipage}\\
\makebox[3em]{$^{41}$}
\begin{minipage}[t]{14cm}
{\it Polytechnic University, Sagamihara, Japan}~$^{I}$

\end{minipage}\\
\makebox[3em]{$^{42}$}
\begin{minipage}[t]{14cm}
{\it Dipartimento di Fisica, Universit\`a 'La Sapienza' and INFN,
Rome, Italy}~$^{B}$

\end{minipage}\\
\makebox[3em]{$^{43}$}
\begin{minipage}[t]{14cm}
{\it Rutherford Appleton Laboratory, Chilton, Didcot, Oxon,
United Kingdom}~$^{D}$

\end{minipage}\\
\makebox[3em]{$^{44}$}
\begin{minipage}[t]{14cm}
{\it Raymond and Beverly Sackler Faculty of Exact Sciences, School of Physics, \\
Tel Aviv University, Tel Aviv, Israel}~$^{Q}$

\end{minipage}\\
\makebox[3em]{$^{45}$}
\begin{minipage}[t]{14cm}
{\it Department of Physics, Tokyo Institute of Technology,
Tokyo, Japan}~$^{I}$

\end{minipage}\\
\makebox[3em]{$^{46}$}
\begin{minipage}[t]{14cm}
{\it Department of Physics, University of Tokyo,
Tokyo, Japan}~$^{I}$

\end{minipage}\\
\makebox[3em]{$^{47}$}
\begin{minipage}[t]{14cm}
{\it Tokyo Metropolitan University, Department of Physics,
Tokyo, Japan}~$^{I}$

\end{minipage}\\
\makebox[3em]{$^{48}$}
\begin{minipage}[t]{14cm}
{\it Universit\`a di Torino and INFN, Torino, Italy}~$^{B}$

\end{minipage}\\
\makebox[3em]{$^{49}$}
\begin{minipage}[t]{14cm}
{\it Universit\`a del Piemonte Orientale, Novara, and INFN, Torino,
Italy}~$^{B}$

\end{minipage}\\
\makebox[3em]{$^{50}$}
\begin{minipage}[t]{14cm}
{\it Department of Physics, University of Toronto, Toronto, Ontario,
Canada M5S 1A7}~$^{M}$

\end{minipage}\\
\makebox[3em]{$^{51}$}
\begin{minipage}[t]{14cm}
{\it Physics and Astronomy Department, University College London,
London, United Kingdom}~$^{D}$

\end{minipage}\\
\makebox[3em]{$^{52}$}
\begin{minipage}[t]{14cm}
{\it Warsaw University, Institute of Experimental Physics,
Warsaw, Poland}

\end{minipage}\\
\makebox[3em]{$^{53}$}
\begin{minipage}[t]{14cm}
{\it Institute for Nuclear Studies, Warsaw, Poland}

\end{minipage}\\
\makebox[3em]{$^{54}$}
\begin{minipage}[t]{14cm}
{\it Department of Particle Physics, Weizmann Institute, Rehovot,
Israel}~$^{R}$

\end{minipage}\\
\makebox[3em]{$^{55}$}
\begin{minipage}[t]{14cm}
{\it Department of Physics, University of Wisconsin, Madison,
Wisconsin 53706, USA}~$^{A}$

\end{minipage}\\
\makebox[3em]{$^{56}$}
\begin{minipage}[t]{14cm}
{\it Department of Physics, York University, Ontario, Canada M3J
1P3}~$^{M}$

\end{minipage}\\
\vspace{30em} \pagebreak[4]


\makebox[3ex]{$^{ A}$}
\begin{minipage}[t]{14cm}
 supported by the US Department of Energy\
\end{minipage}\\
\makebox[3ex]{$^{ B}$}
\begin{minipage}[t]{14cm}
 supported by the Italian National Institute for Nuclear Physics (INFN) \
\end{minipage}\\
\makebox[3ex]{$^{ C}$}
\begin{minipage}[t]{14cm}
 supported by the German Federal Ministry for Education and Research (BMBF), under
 contract Nos. 05 HZ6PDA, 05 HZ6GUA, 05 HZ6VFA and 05 HZ4KHA\
\end{minipage}\\
\makebox[3ex]{$^{ D}$}
\begin{minipage}[t]{14cm}
 supported by the Science and Technology Facilities Council, UK\
\end{minipage}\\
\makebox[3ex]{$^{ E}$}
\begin{minipage}[t]{14cm}
 supported by an FRGS grant from the Malaysian government\
\end{minipage}\\
\makebox[3ex]{$^{ F}$}
\begin{minipage}[t]{14cm}
 supported by the US National Science Foundation. Any opinion,
 findings and conclusions or recommendations expressed in this material
 are those of the authors and do not necessarily reflect the views of the
 National Science Foundation.\
\end{minipage}\\
\makebox[3ex]{$^{ G}$}
\begin{minipage}[t]{14cm}
 supported by the Polish Ministry of Science and Higher Education as a scientific project No.
 DPN/N188/DESY/2009\
\end{minipage}\\
\makebox[3ex]{$^{ H}$}
\begin{minipage}[t]{14cm}
 supported by the Polish Ministry of Science and Higher Education
 as a scientific project (2009-2010)\
\end{minipage}\\
\makebox[3ex]{$^{ I}$}
\begin{minipage}[t]{14cm}
 supported by the Japanese Ministry of Education, Culture, Sports, Science and Technology
 (MEXT) and its grants for Scientific Research\
\end{minipage}\\
\makebox[3ex]{$^{ J}$}
\begin{minipage}[t]{14cm}
 supported by the Korean Ministry of Education and Korea Science and Engineering
 Foundation\
\end{minipage}\\
\makebox[3ex]{$^{ K}$}
\begin{minipage}[t]{14cm}
 supported by FNRS and its associated funds (IISN and FRIA) and by an Inter-University
 Attraction Poles Programme subsidised by the Belgian Federal Science Policy Office\
\end{minipage}\\
\makebox[3ex]{$^{ L}$}
\begin{minipage}[t]{14cm}
 supported by the Spanish Ministry of Education and Science through funds provided by
 CICYT\
\end{minipage}\\
\makebox[3ex]{$^{ M}$}
\begin{minipage}[t]{14cm}
 supported by the Natural Sciences and Engineering Research Council of Canada (NSERC) \
\end{minipage}\\
\makebox[3ex]{$^{ N}$}
\begin{minipage}[t]{14cm}
 partially supported by the German Federal Ministry for Education and Research (BMBF)\
\end{minipage}\\
\makebox[3ex]{$^{ O}$}
\begin{minipage}[t]{14cm}
 supported by RF Presidential grant N 41-42.2010.2 for the Leading
 Scientific Schools and by the Russian Ministry of Education and Science through its
 grant for Scientific Research on High Energy Physics\
\end{minipage}\\
\makebox[3ex]{$^{ P}$}
\begin{minipage}[t]{14cm}
 supported by the Netherlands Foundation for Research on Matter (FOM)\
\end{minipage}\\
\makebox[3ex]{$^{ Q}$}
\begin{minipage}[t]{14cm}
 supported by the Israel Science Foundation\
\end{minipage}\\
\makebox[3ex]{$^{ R}$}
\begin{minipage}[t]{14cm}
 supported in part by the MINERVA Gesellschaft f\"ur Forschung GmbH, the Israel Science
 Foundation (grant No. 293/02-11.2) and the US-Israel Binational Science Foundation \
\end{minipage}\\
\vspace{30em} \pagebreak[4]


\makebox[3ex]{$^{ a}$}
\begin{minipage}[t]{14cm}
also affiliated with University College London,
 United Kingdom\
\end{minipage}\\
\makebox[3ex]{$^{ b}$}
\begin{minipage}[t]{14cm}
now at University of Salerno, Italy\
\end{minipage}\\
\makebox[3ex]{$^{ c}$}
\begin{minipage}[t]{14cm}
now at Queen Mary University of London, United Kingdom\
\end{minipage}\\
\makebox[3ex]{$^{ d}$}
\begin{minipage}[t]{14cm}
also working at Max Planck Institute, Munich, Germany\
\end{minipage}\\
\makebox[3ex]{$^{ e}$}
\begin{minipage}[t]{14cm}
also Senior Alexander von Humboldt Research Fellow at Hamburg University,
 Institute of Experimental Physics, Hamburg, Germany\
\end{minipage}\\
\makebox[3ex]{$^{ f}$}
\begin{minipage}[t]{14cm}
also at Cracow University of Technology, Faculty of Physics,
 Mathemathics and Applied Computer Science, Poland\
\end{minipage}\\
\makebox[3ex]{$^{ g}$}
\begin{minipage}[t]{14cm}
supported by the research grant No. 1 P03B 04529 (2005-2008)\
\end{minipage}\\
\makebox[3ex]{$^{ h}$}
\begin{minipage}[t]{14cm}
now at Rockefeller University, New York, NY
 10065, USA\
\end{minipage}\\
\makebox[3ex]{$^{ i}$}
\begin{minipage}[t]{14cm}
now at DESY group FS-CFEL-1\
\end{minipage}\\
\makebox[3ex]{$^{ j}$}
\begin{minipage}[t]{14cm}
now at Institute of High Energy Physics, Beijing,
 China\
\end{minipage}\\
\makebox[3ex]{$^{ k}$}
\begin{minipage}[t]{14cm}
now at DESY group FEB, Hamburg, Germany\
\end{minipage}\\
\makebox[3ex]{$^{ l}$}
\begin{minipage}[t]{14cm}
also at Moscow State University, Russia\
\end{minipage}\\
\makebox[3ex]{$^{ m}$}
\begin{minipage}[t]{14cm}
now at University of Liverpool, United Kingdom\
\end{minipage}\\
\makebox[3ex]{$^{ n}$}
\begin{minipage}[t]{14cm}
on leave of absence at CERN, Geneva, Switzerland\
\end{minipage}\\
\makebox[3ex]{$^{ o}$}
\begin{minipage}[t]{14cm}
now at CERN, Geneva, Switzerland\
\end{minipage}\\
\makebox[3ex]{$^{ p}$}
\begin{minipage}[t]{14cm}
now at Goldman Sachs, London, UK\
\end{minipage}\\
\makebox[3ex]{$^{ q}$}
\begin{minipage}[t]{14cm}
also at Institute of Theoretical and Experimental Physics, Moscow, Russia\
\end{minipage}\\
\makebox[3ex]{$^{ r}$}
\begin{minipage}[t]{14cm}
also at INP, Cracow, Poland\
\end{minipage}\\
\makebox[3ex]{$^{ s}$}
\begin{minipage}[t]{14cm}
also at FPACS, AGH-UST, Cracow, Poland\
\end{minipage}\\
\makebox[3ex]{$^{ t}$}
\begin{minipage}[t]{14cm}
partially supported by Warsaw University, Poland\
\end{minipage}\\
\makebox[3ex]{$^{ u}$}
\begin{minipage}[t]{14cm}
partially supported by Moscow State University, Russia\
\end{minipage}\\
\makebox[3ex]{$^{ v}$}
\begin{minipage}[t]{14cm}
also affiliated with DESY, Germany\
\end{minipage}\\
\makebox[3ex]{$^{ w}$}
\begin{minipage}[t]{14cm}
now at Japan Synchrotron Radiation Research Institute (JASRI), Hyogo, Japan\
\end{minipage}\\
\makebox[3ex]{$^{ x}$}
\begin{minipage}[t]{14cm}
also at University of Tokyo, Japan\
\end{minipage}\\
\makebox[3ex]{$^{ y}$}
\begin{minipage}[t]{14cm}
now at Kobe University, Japan\
\end{minipage}\\
\makebox[3ex]{$^{ z}$}
\begin{minipage}[t]{14cm}
supported by DESY, Germany\
\end{minipage}\\
\makebox[3ex]{$^{\dagger}$}
\begin{minipage}[t]{14cm}
 deceased \
\end{minipage}\\
\makebox[3ex]{$^{aa}$}
\begin{minipage}[t]{14cm}
STFC Advanced Fellow\
\end{minipage}\\
\makebox[3ex]{$^{ab}$}
\begin{minipage}[t]{14cm}
nee Korcsak-Gorzo\
\end{minipage}\\
\makebox[3ex]{$^{ac}$}
\begin{minipage}[t]{14cm}
This material was based on work supported by the
 National Science Foundation, while working at the Foundation.\
\end{minipage}\\
\makebox[3ex]{$^{ad}$}
\begin{minipage}[t]{14cm}
also at Max Planck Institute, Munich, Germany, Alexander von Humboldt
 Research Award\
\end{minipage}\\
\makebox[3ex]{$^{ae}$}
\begin{minipage}[t]{14cm}
now at Nihon Institute of Medical Science, Japan\
\end{minipage}\\
\makebox[3ex]{$^{af}$}
\begin{minipage}[t]{14cm}
now at SunMelx Co. Ltd., Tokyo, Japan\
\end{minipage}\\
\makebox[3ex]{$^{ag}$}
\begin{minipage}[t]{14cm}
now at Osaka University, Osaka, Japan\
\end{minipage}\\
\makebox[3ex]{$^{ah}$}
\begin{minipage}[t]{14cm}
now at University of Bonn, Germany\
\end{minipage}\\
\makebox[3ex]{$^{ai}$}
\begin{minipage}[t]{14cm}
also at \L\'{o}d\'{z} University, Poland\
\end{minipage}\\
\makebox[3ex]{$^{aj}$}
\begin{minipage}[t]{14cm}
member of \L\'{o}d\'{z} University, Poland\
\end{minipage}\\
\makebox[3ex]{$^{ak}$}
\begin{minipage}[t]{14cm}
now at Lund University, Lund, Sweden\
\end{minipage}\\
\makebox[3ex]{$^{al}$}
\begin{minipage}[t]{14cm}
also at University of Podlasie, Siedlce, Poland\
\end{minipage}\\

}


\pagenumbering{arabic} 
\pagestyle{plain}
\section{Introduction}
\label{sec-int}

Heavy-quark production in $ep$ interactions in deep inelastic scattering (DIS) is dominated by the Boson Gluon Fusion (BGF) process. Heavy-quark production provides a two-fold test of perturbative quantum chromodynamics (pQCD); a study of the BGF process and the higher-order corrections to it, and an independent check of the validity of the gluon density in the proton extracted from the inclusive DIS data. Of the two heavy quarks whose production is accessible at HERA, $c$ and $b$, the latter is strongly suppressed due to its smaller electric charge and larger mass.

The production of charm via the identification of $D$ and $D^{*}$ mesons in DIS has been extensively studied at HERA in the kinematic range $1 < Q^{2} < 1000\gev^{2}$, $p_{T}(D, D^{*}) > 1.5\gev$ ~\cite{pl:b407:402, np:b545:21, epj:c12:35, pl:b528:199, pr:d69:012004, Aktas:2004ka, Chekanov:2007ch, Chekanov:2008yd}, where $Q^{2}$ is the negative squared four-momentum exchange at the electron vertex and $p_{T}$ is the transverse momentum. The results are consistent with the calculations of pQCD. The fragmentation fraction $f(c\to \Lambda_{c}^{+})$ has been measured by the ZEUS collaboration in the photoproduction regime~\cite{Chekanov:2005mm}. The obtained fragmentation fraction is larger than but consistent within uncertainties with the average from $e^{+}e^{-}$ collisions~\cite{Gladilin:1999pj}.

In this paper, a charm quark in the final state was identified by the presence of a charmed hadron. The production of $D^{+}$ mesons and $\Lambda_{c}^{+}$ baryons was studied using the decays\footnote{The charge conjugated modes are implied throughout this paper.} $D^{+}\to K^{0}_{S} \pi^{+}$, $\Lambda_{c}^{+}\to p K^{0}_{S}$ and $\Lambda_{c}^{+}\to \Lambda \pi^{+}$. These decay channels were chosen since the presence of a neutral strange hadron in the final state significantly reduces the combinatorial background. Measurements of $D^{+}$ and $\Lambda_{c}^{+}$ cross sections provide information about both $c$-quark production and its fragmentation.

With respect to previous studies, in this analysis the kinematic region of the measurement is extended to very low transverse momenta of the produced charmed hadrons. No explicit cut on the transverse momenta of the reconstructed charmed hadrons was applied. This is particularly relevant at low $Q^{2}$, where charm quarks are predominantly produced with low transverse momentum. In addition, $\Lambda_{c}^{+}$ production was studied for the first time at HERA in DIS. From a comparison of the $D^{+}$ and $\Lambda_{c}^{+}$ cross sections, the fragmentation fraction $f(c\to \Lambda_{c}^{+})$ is extracted.

\section{Experimental set-up}
\label{sec-exp}

The analysis was performed with data taken from 1996 to 2000 corresponding to a luminosity of $120.4 \pm 2.4\pbi$. The sample consists of $38.6\pbi$ of $e^{+}p$ data collected at a centre-of-mass energy of $300\gev$ and of $65.1\pbi$ collected at $318\gev$, plus $16.7\pbi$ of $e^{-}p$ data collected at $318\gev$.\footnote{Hereafter, both electrons and positrons are referred to as electrons, unless explicitly stated otherwise.}

\Zdetdesc

\Zctddesc\ZcoosysfnBeta The transverse-momentum resolution for full-length tracks was $\sigma(p_T)/p_T=0.0058p_T\oplus0.0065\oplus0.0014/p_T$, with $p_T$ in $\Gev$.

To estimate the ionisation energy loss per unit length, $dE/dx$, of particles in the CTD\cite{pl:b481:213,*epj:c18:625,*thesis:bartsch:2007,*Chekanov:2008aaa}, the truncated mean of the anode-wire pulse heights was calculated, which removes the $10\%$ lowest and at least the $30\%$ highest pulses depending on the number of saturated hits. The measured $dE/dx$ values were corrected by normalising to the measured average $dE/dx$ for tracks around the region of minimum ionisation for pions with momentum $p$ satisfying $0.3~<~p~<~0.4\gev$. Henceforth, $dE/dx$ is quoted in units of minimum ionising particles (mips).

\Zcaldesc

The position of the scattered electron at the CAL was determined by combining information from the CAL and, where available, the small-angle rear tracking detector (SRTD)~\cite{nim:a401:63} and the hadron-electron separator (HES)~\cite{nim:a277:176}.

The luminosity was measured from the rate of the bremsstrahlung process $ep\to e\gamma p$, where the photon was measured in a lead--scintillator calorimeter \cite{desy-92-066,*zfp:c63:391,*acpp:b32:2025} placed in the HERA tunnel at $Z=-107\met$.

\section{Theoretical predictions}
\label{sec-theo}

The next-to-leading-order (NLO) QCD predictions for the $c\bar{c}$ production cross sections were obtained using the HVQDIS program~\cite{pr:d57:2806} based on the fixed-flavour-number scheme (FFNS). In this scheme, only light quarks ($u$, $d$ and $s$) and gluons are included in the proton parton density functions (PDFs) which obey the DGLAP equations~\cite{sovjnp:15:438,*sovjnp:20:94,*np:b126:298,*jetp:46:641}, and the $c\bar{c}$ pair is produced via the BGF mechanism~\cite{np:b452:109,*pl:b353:535} with NLO corrections~\cite{np:b392:162,*np:b392:229}. The presence of different large scales, $Q$, $p_{T}$ and the mass of the $c$ quark, $m_{c}$, can spoil the convergence of the perturbative series because the neglected terms of orders higher than $\alpha_{s}^{2}$ (where $\alpha_{s}$ is the strong coupling constant) contain $\log(Q^{2}/m_{c}^{2})$ factors which can become large. The FFNS variant of the ZEUS-S NLO QCD fit~\cite{pr:d67:012007,*misc:www:zeus2002} to structure function data was used as the parametrisation of the proton PDFs. In this fit, $\alpha_{s}(M_{Z})$ was set to $0.118$ and the mass of the charm quark was set to $1.5\gev$; the same mass was used in the HVQDIS calculation. The renormalisation and factorisation scales were set to $\mu_{R} = \mu_{F} = \sqrt{Q^{2} + 4m_{c}^{2}}$. The charm fragmentation to the $D^{+}$ meson was modelled using the Peterson function~\cite{pr:d27:105} with the Peterson parameter, $\epsilon$, set to $0.079$~\cite{Chekanov:2008ur}. For the hadronisation fraction, $f(c\to D^{+})$, the value $0.216^{+0.021}_{-0.029}$ was used~\cite{Chekanov:2007ch}.

The HVQDIS predictions for the production of $D^{+}$ mesons are affected by the theoretical uncertainties listed below. The uncertainty on the total cross section is given in parentheses:

\begin{itemize}

\item
the ZEUS PDF uncertainties were propagated from the experimental uncertainties of the fitted data ($^{+5.3\%}_{-5.2\%}$);

\item
the charm quark mass was changed consistently in the PDF fit and in HVQDIS by $\pm0.15\gev$($^{+15.2\%}_{-13.5\%}$);

\item
the renormalisation scale was varied by a factor 2 ($^{+19.7\%}_{-12.6\%}$);

\item
the factorisation scale was changed by a factor 2 independently of the renormalisation scale ($^{+13.1\%}_{-21.7\%}$);

\item
the $\epsilon$ parameter of the Peterson fragmentation function was changed to 0.01 and 0.1~\cite{Chekanov:2008ur,Aaron:2008tt}. This modification affects the shapes of the $p_{T}$, $Q^{2}$ and $x$ distributions ($^{+0.1\%}_{-0.4\%}$).

\end{itemize}

\section{Monte Carlo models}
\label{sec-mc}

The detector acceptance was modelled using the {\sc Rapgap 3.00}~\cite{cpc:86:147} Monte Carlo (MC) program, interfaced with {\sc Heracles 4.6.1}~\cite{cpc:69:155} in order to incorporate first-order electroweak corrections. The generated events were passed through a full simulation of the detector, using {\sc Geant 3.13}~\cite{tech:cern-dd-ee-84-1}, and finally processed and selected in the same way as the data.

The MC was used to simulate events containing charm quarks produced in the BGF process. The {\sc Rapgap} generator used leading-order matrix elements with leading-logarithmic parton showers. The CTEQ5L~\cite{epj:c12:375} PDFs were used for the proton. The charm-quark mass was set to $1.5\gev$. Charm fragmentation was simulated using the Lund string model~\cite{Andersson:1983ia}. The $D^{+}$ and $\Lambda_{c}^{+}$ hadrons originating from beauty decays were accounted for by including a {\sc Rapgap} $b$-quark sample where the $b$-quark mass was set to $4.75\gev$. An additional sample where charm was produced by the process $cg\to cg$ was generated and was used to study the model dependence of the simulation. For this process, the charm quark was treated as a part of the structure of the photon. The processes $gg\to c\bar{c}$ and $q\bar{q}\to c\bar{c}$ were not included because their contribution estimated using the {\sc Rapgap} MC was found to be less than $1\%$ in the studied kinematic range.

In general, the MC gives a reasonable description of the data for DIS and $D^{+}$-meson variables when compared at detector level. To improve the description further, {\sc Rapgap} was reweighted to reproduce the $p_{T}(D^{+})$ distribution observed in the data. The same weights used for $D^{+}$ mesons were also applied to $D_{s}^{+}$ and $\Lambda_{c}^{+}$ hadrons.

\section{Kinematic reconstruction and event selection}
\label{sec-reco}

A three-level trigger system was used to select events online~\cite{zeus:1993:bluebook,uproc:chep:1992:222}. At the third level, an electron with an energy greater than $4\gev$ and a position outside a box of $24\times 12\cm^{2}$ centred around the beampipe on the face of the rear calorimeter was required by a fully inclusive DIS trigger which had a high acceptance for $Q^{2} \gtrsim 1\gev^{2}$. However, this trigger was heavily prescaled and the equivalent luminosity is $17\pbi$.

Additionally, events above $Q^{2} \approx 20\gev^{2}$ were selected by a medium-$Q^{2}$ trigger. The only difference to the inclusive DIS trigger is that the position of the scattered electron on the RCAL face had to lie outside a circle centred around the beampipe of radius between $25$ and $35\cm$, depending on the running period.

The fraction of the electron energy transferred to the proton in its rest frame, $y$, as well as the kinematic variables $Q^{2}$ and Bjorken $x$, were reconstructed offline using the electron method~\cite{proc:hera:1991:23,*hoeger} (denoted by the subscript $e$), which uses the energy and angle of the scattered electron. The inelasticity $y$ was also obtained using the Jacquet-Blondel (JB) method~\cite{proc:epfacility:1979:391}. The double angle (DA) method~\cite{proc:hera:1991:23,*hoeger}, which relies on the angles of the scattered electron and the hadronic-energy flow, was used as a systematic check.

The following requirements were imposed offline:

\begin{itemize}

\item
$38 \, < \,  \delta \,  < \, 65\gev$, where $\delta = \sum E_i(1-\cos\theta_i)$ and $E_{i}$ and $\theta_{i}$ are the energy and the polar angle of the $i^{th}$ energy-flow object (EFO)~\cite{thesis:briskin:1998} reconstructed from charged tracks, as measured in the CTD, and energy clusters measured in the CAL. The sum $i$ runs over all EFOs~\cite{pl:b303:183};

\item
$E_{e}^{'} > 10\gev$, where $E_{e}^{'}$ is the energy of the scattered electron identified using a neural-network algorithm~\cite{nim:a365:508,nim:a391:360};

\item
$E_{{\rm cone}} < 5\gev$, where $E_{{\rm cone}}$ is the calorimeter energy measured in a cone around the electron position that was not assigned to the electron cluster. The cone was defined by $R_{{\rm cone}} < 0.8$ with $R_{{\rm cone}} = \sqrt{(\Delta \phi)^{2} + (\Delta \eta)^{2}}$;

\item
a match between the tracking and the calorimeter information for electrons well within the CTD acceptance, $17^{\circ} < \theta_{e} < 149^{\circ}$. For $\theta_{e}$ outside this region, the cut $\delta > 44\gev$ was imposed;

\item
for events with the scattered electron reconstructed within the SRTD acceptance, the impact position of the electron on the face of the RCAL had to be outside the region $26 \times 14\cm^{2}$ centred on $X = Y = 0$. If the electron position was reconstructed without using SRTD information, a box cut of $26 \times 20\cm^{2}$ was imposed;

\item
$1.5 < Q^{2}_{e} < 1000\gev^{2}$;

\item
$y_{{\rm JB}} > 0.02$ and $y_{e} < 0.7$;

\item
a primary vertex position in the range $|Z_{\textnormal{vertex}}| < 50\cm$.

\end{itemize}

This analysis used charged tracks measured in the CTD that were assigned either to the primary or to a secondary vertex. The tracks were required to have transverse momenta $p_{T} > 0.15\gev$ and pseudorapidity in the laboratory frame $|\eta| < 1.75$, restricting the study to a region where the CTD track acceptance and resolution were high. Candidates for long-lived neutral strange hadrons decaying to two charged particles were identified by selecting pairs of oppositely charged tracks, fitted to a displaced secondary vertex. The events were required to have at least one such candidate.

\section{Strange-particle reconstruction}
\label{sec-strange-reco}

The $K^{0}_{S}$ mesons were identified by their charged decay mode, $K^{0}_{S}\to \pi^{+} \pi^{-}$. Both tracks were assigned the mass of the charged pion and the invariant mass, $M(\pi^{+} \pi^{-})$, of each track pair was calculated. Additional requirements to select $K^{0}_{S}$ were imposed:

\begin{itemize}

\item
$M(e^{+}e^{-}) > 50\mev$, where the electron mass was assigned to each track, to eliminate tracks from photon conversions;

\item
$M(p\pi) > 1121\mev$, where the proton mass was assigned to the track with higher momentum, to eliminate $\Lambda$ contamination in the $K^{0}_{S}$ signal;

\item
$\cos\theta_{XY} > 0.98$, where $\theta_{XY}$ is defined as the angle between the momentum vector of the $K^{0}_{S}$ candidate and the vector defined by the primary interaction vertex and the $K^{0}_{S}$ decay vertex in the $X$-$Y$ plane;

\item
$483 < M(\pi^{+} \pi^{-}) < 513\mev$;

\item
$|\eta(K^{0}_{S})| < 1.6$.

\end{itemize}

The $\Lambda$ candidates were reconstructed by their charged decay mode to $p\pi^{-}$. The track with the larger momentum was assigned the mass of the proton, while the other was assigned the mass of the charged pion, as the decay proton always has a larger momentum than the pion, provided the $\Lambda$ momentum is greater than $0.3\gev$. Additional requirements to select $\Lambda$ were imposed:

\begin{itemize}

\item
$M(e^{+}e^{-}) > 50\mev$;

\item
$M(\pi^{+}\pi^{-}) < 483\mev$, where the charged pion mass was assigned to both tracks, to remove $K^{0}_{S}$ contamination in the $\Lambda$ signal;

\item
$\cos\theta_{XY} > 0.98$;

\item
$1112 < M(p\pi) < 1121\mev$;

\item
$|\eta(\Lambda)| < 1.6$.

\end{itemize}

\begin{figure}[hbtp]
\centerline{\epsfig{file=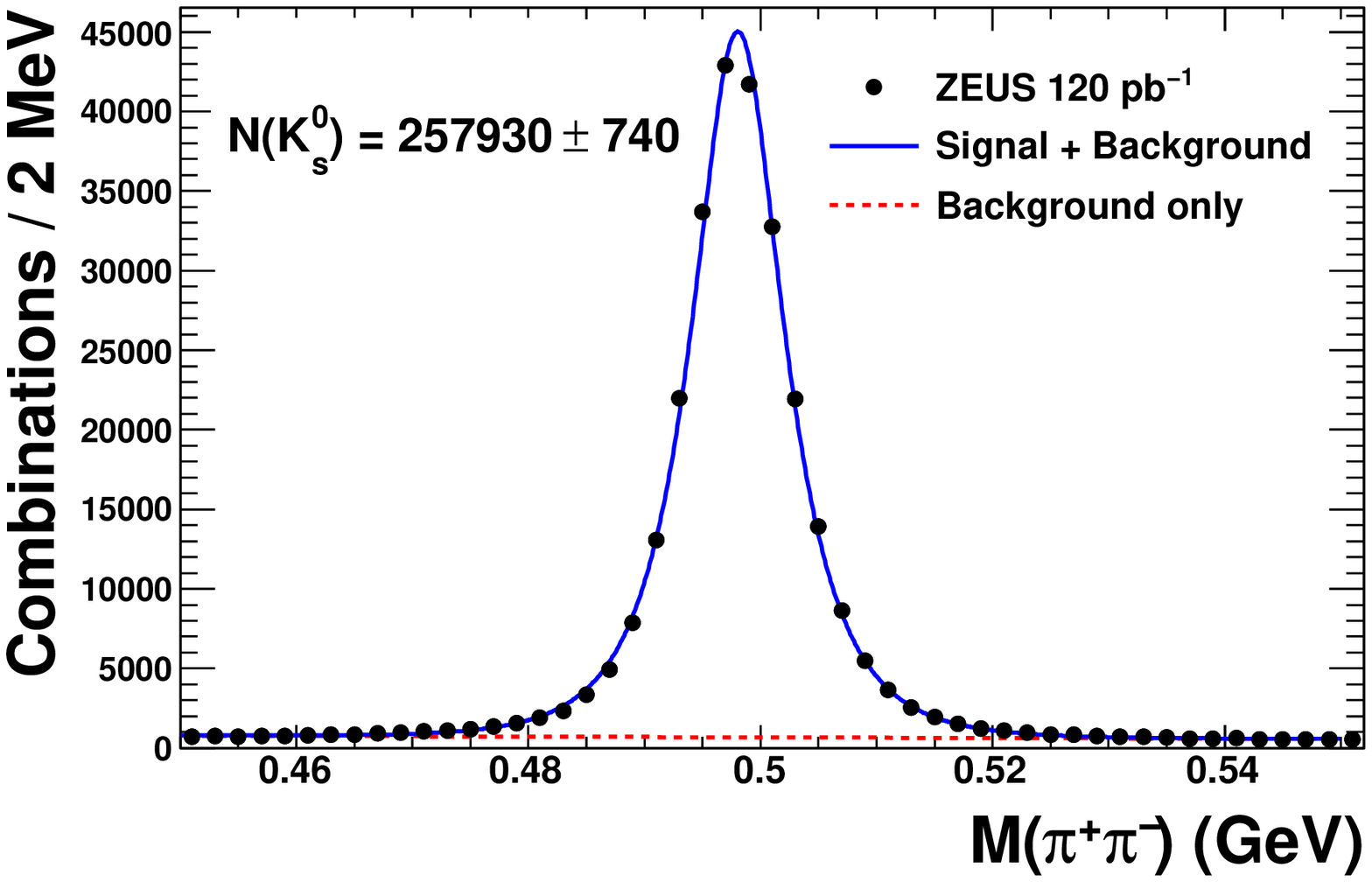,height=6cm}\put(-165,158){\Huge\bfseries ZEUS}\put(-70,100){\makebox(0,0)[tl]{\large (a)}}}
\centerline{\epsfig{file=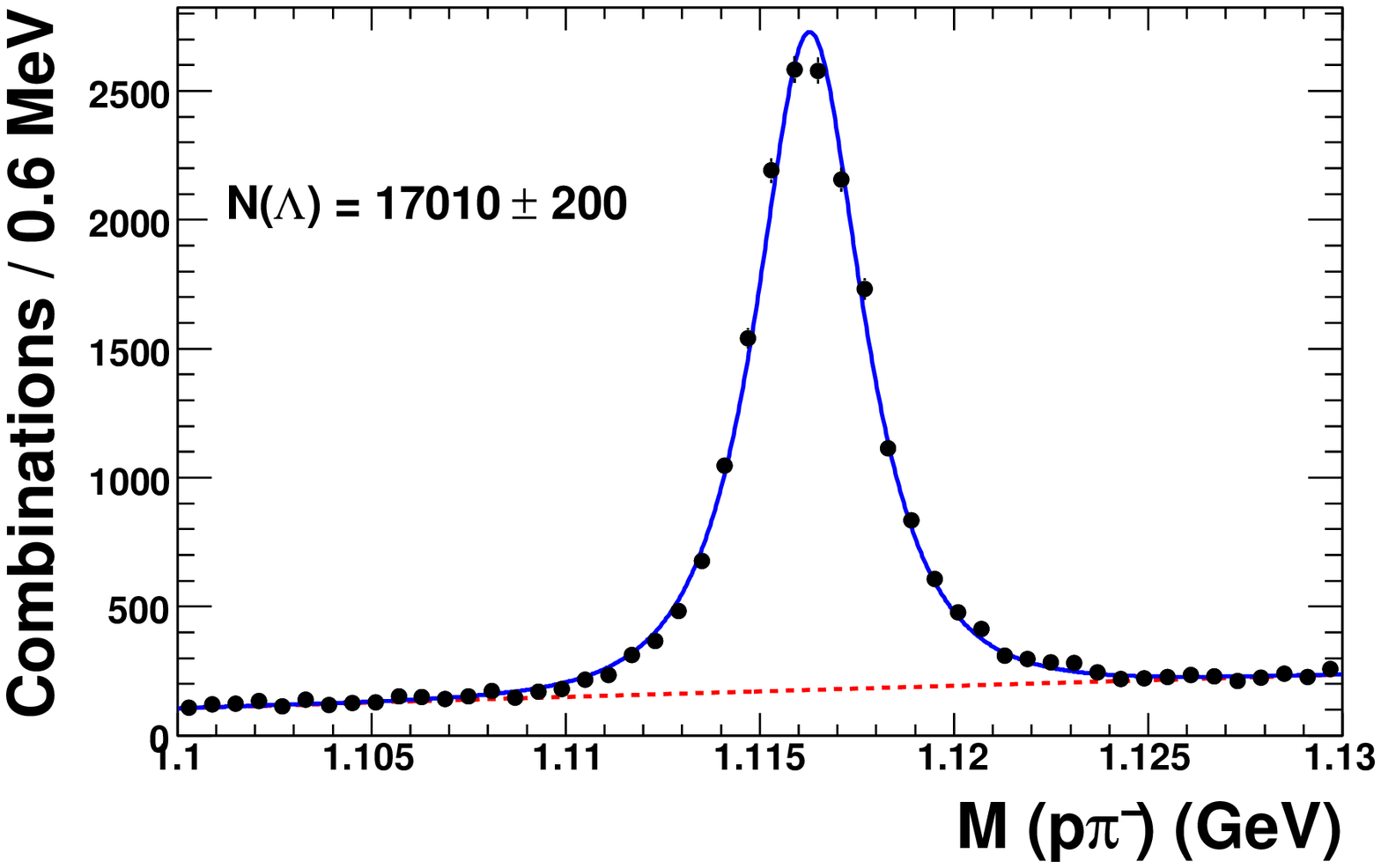,height=6cm}\put(-70,100){\makebox(0,0)[tl]{\large (b)}}}
\centerline{\epsfig{file=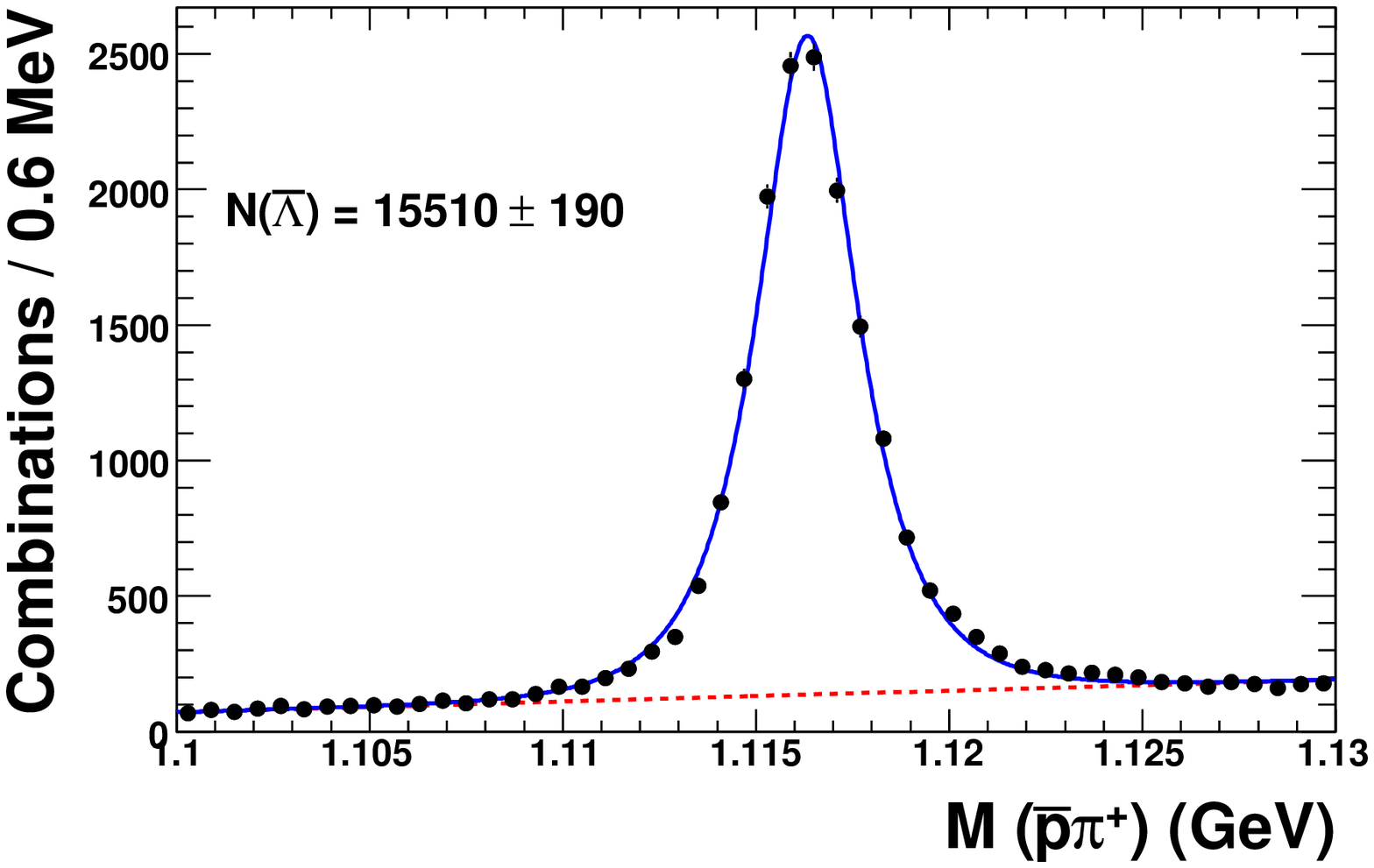,height=6cm}\put(-70,100){\makebox(0,0)[tl]{\large (c)}}}
\caption{Mass distributions of the secondary vertex candidates in the (a) $K^{0}_{S}$, (b) $\Lambda$ and (c) $\bar{\Lambda}$ samples. The statistical uncertainties are in general smaller than the point size. For illustration the data have been fitted using the sum of a ``modified'' Gaussian function~\protect\cite{Chekanov:2005mm} and a linear background.}
\label{fig:peaks_v0}
\end{figure}

Figure~\ref{fig:peaks_v0} shows the invariant-mass spectra of $K^{0}_{S}$, $\Lambda$ and $\bar{\Lambda}$ candidates. Distributions of the reconstructed proper lifetime for these particles based on the same data sample as analysed in this paper were found to be satisfactory~\cite{Chekanov:2006wz}.

\section{Reconstruction of charmed hadrons}
\label{sec-charm-reco}

The production of $D^{+}$ and $\Lambda_{c}^{+}$ hadrons was measured in the range of transverse momentum $0 < p_{T}(D^{+},\Lambda_{c}^{+}) < 10\gev$ and pseudorapidity $|\eta(D^{+},\Lambda_{c}^{+})| < 1.6$. Strange-hadron candidates were combined with a further track measured in the CTD which was assigned to the primary interaction vertex. The combinatorial background was significantly reduced by requiring $p_{T}(D^{+})/E_{T}^{\theta>10^{\circ}} > 0.1$ and $p_{T}(\Lambda_{c}^{+})/E_{T}^{\theta>10^{\circ}} > 0.12$, where the transverse energy $E_{T}^{\theta>10^{\circ}}$ was evaluated as $E_{T}^{\theta>10^{\circ}} = \sum_{i,\theta_{i} > 10^{\circ}} (E_{i}\sin{\theta_{i}})$. The sum runs over all energy deposits in the CAL with a polar angle $\theta$ above $10^{\circ}$. The details of the reconstruction of the three different decay channels are given in the next subsections.

\boldmath
\subsection{Reconstruction of the decay $D^{+}\to K^{0}_{S} \pi^{+}$}
\unboldmath

The $D^{+}$ mesons were reconstructed from the decay channel $D^{+}\to K^{0}_{S} \pi^{+}$. In each event, $D^{+}$ candidates were formed from combinations of $K^{0}_{S}$ candidates reconstructed as described in Section~\ref{sec-strange-reco} with further tracks assumed to be pions. The pion candidates were required to have $p_{T}(\pi^{+})/E_{T}^{\theta>10^{\circ}} > 0.04$. Only pion candidates with $dE/dx < 1.5$~mips were considered. Further reduction of the combinatorial background was achieved by cutting on the angle between the pion in the $D^{+}$ rest frame and the $D^{+}$ flight direction, $\theta^{*}(\pi^{+})$. Different cuts depending on $p_{T}(D^{+})$ were used to ensure optimal background suppression:

\begin{itemize}

\item
$\cos\theta^{*}(\pi^{+}) < 0.9$ \;\;\; for \;\;\; $0.0 < p_{T}(D^{+}) < 1.5\gev;$

\item
$\cos\theta^{*}(\pi^{+}) < 0.8$ \;\;\; for \;\;\; $1.5 < p_{T}(D^{+}) < 3.0\gev;$

\item
$\cos\theta^{*}(\pi^{+}) < 0.6$ \;\;\; for \;\;\; $3.0 < p_{T}(D^{+}) < 10.0\gev.$

\end{itemize}

The $K^{0}_{S}\pi^{+}$ invariant-mass distribution was fitted with the sum of contributions from the signal, the non-resonant background and a reflection caused by $D_{s}^{+}\to K^{0}_{S}K^{+}$ decays. The signal was described by a Gaussian function defined as:
\begin{equation}
g(\sigma,M_{0};m) = \frac{1}{\sqrt{2\pi}\sigma} \exp{\frac{-(m-M_{0})^{2}}{2\sigma^{2}}},
\end{equation}
where $M_{0}$ and $\sigma$ are the resonance mass and width, respectively. For the background a sum of Chebyshev polynomials up to the second order was used:
\begin{equation}
b(A,B,C;y(m)) = A \cdot (1 + B \cdot y + C \cdot (2y^{2} - 1)),
\label{background-cheby}
\end{equation}
where $y(m) = (2m - m_{{\rm max}} - m_{{\rm min}}) \, / \, (m_{{\rm max}} - m_{{\rm min}})$ and $m_{{\rm max}} (m_{{\rm min}}) = 2.1 (1.6)\gev$ is the upper (lower) limit of the fitted range.

The mass distribution of the reflection $r(m)$ caused by the decay $D_{s}^{+}\to K^{0}_{S}K^{+}\to \pi^{+}\pi^{-}K^{+}$ was obtained from $D_{s}^{+}$ combinations in the Monte Carlo at detector level matched to the same decay at generator level. The normalisation of the reflection with respect to the Gaussian signal assumed for $D^{+}\to K^{0}_{S}\pi^{+}$ decays is based on previously measured fragmentation fractions $f$~\cite{Chekanov:2007ch} and branching ratios $\mathcal{B}$~\cite{Amsler:2008zzb} (see also Table~\ref{tab-branching-ratios}) and the detector acceptances for both decay channels. For this purpose, the invariant mass distribution of the reflection was normalised to unity and then multiplied by the expected ratio of $D_{s}^{+}$ to $D^{+}$ mesons:
\begin{equation}
R = \frac{f(c\to D_{s}^{+}) \cdot \mathcal{B}(D_{s}^{+}\to K^{0}_{S}K^{+}\to \pi^{+}\pi^{-}K^{+})}{f(c\to D^{+}) \cdot \mathcal{B}(D^{+}\to K^{0}_{S}\pi^{+}\to \pi^{+}\pi^{-}\pi^{+})} \cdot \frac{\mathcal{A}(D_{s}^{+})}{\mathcal{A}(D^{+})} = 0.44 \pm 0.10,
\label{ratio_reflection}
\end{equation}
where $\mathcal{A}(D_{s}^{+})$ and $\mathcal{A}(D^{+})$ are the reconstruction acceptances for $D_{s}^{+}$ and $D^{+}$ mesons, respectively, as obtained from the Monte Carlo. The resulting fitting function is given by:
\begin{equation}
F(A,B,C,D,\sigma,M_{0};m) = b(A,B,C;y(m)) + D \cdot [r(m) + g(\sigma,M_{0};m)],
\end{equation}
where the parameters $A$, $B$, $C$, $D$, $\sigma$ and $M_{0}$ were determined by the fit.

\begin{figure}[hbtp]
\epsfysize=14cm
\centerline{\mbox{\epsfig{file=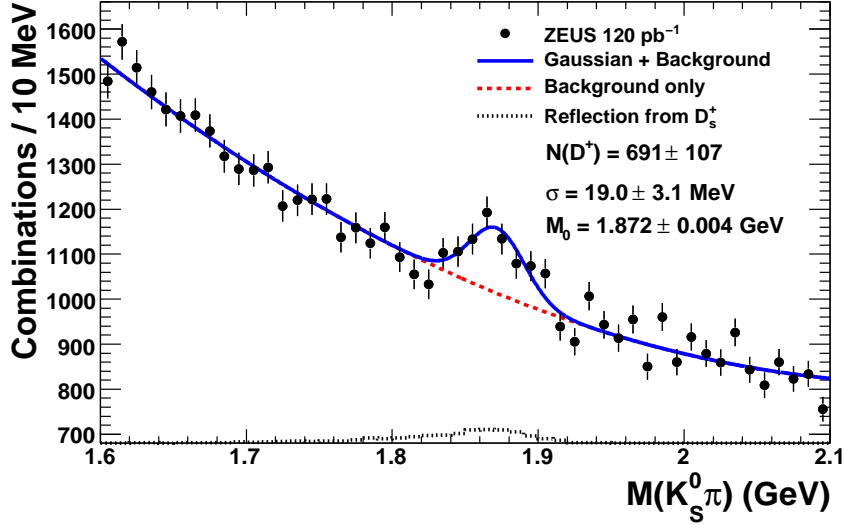,width=12cm}}\put(-210,202){\Huge\bfseries ZEUS}}
\caption{The $M(K^{0}_{S}\pi^{+})$ distribution (dots) for $D^{+}$ candidates. The reflection caused by the decay $D_{s}^{+}\to K^{0}_{S}K^{+}$ has been subtracted as described in the text. The solid curve represents a fit to the sum of a Gaussian signal and a background function, while the background contribution alone is given by the dashed curve. The dotted histogram shows the reflection scaled as described in the text with an offset of 680 to position it at the bottom of the figure.}
\label{fig:peak_dpm}
\end{figure}

Figure~\ref{fig:peak_dpm} shows the invariant mass spectrum for the $D^{+}$ candidates after the reflection was subtracted using the fit, resulting in a 20\% reduction in the number of $D^{+}$ mesons. A clear signal is visible. The fit yielded a $D^{+}$ mass of $1872 \pm 4\mev$, in agreement with the PDG value~\cite{Amsler:2008zzb}. The width of the signal was $19.0 \pm 3.1\mev$, reflecting the detector resolution. The number of $D^{+}$ mesons yielded by the fit was $N(D^{+}) = 691 \pm 107$.

In order to extract the $D^{+}$-meson yields in bins of $p_{T}^{2}(D^{+})$, $\eta(D^{+})$, $Q^{2}$ and $x$, the signals in all analysis bins of a given quantity were fitted simultaneously, fixing the ratios of the widths in the bins to the Monte Carlo prediction. All other parameters including the masses were left free for all bins in the simultaneous fit.

\begin{figure}[hbtp]
\epsfysize=14cm
\centerline{\mbox{\epsfig{file=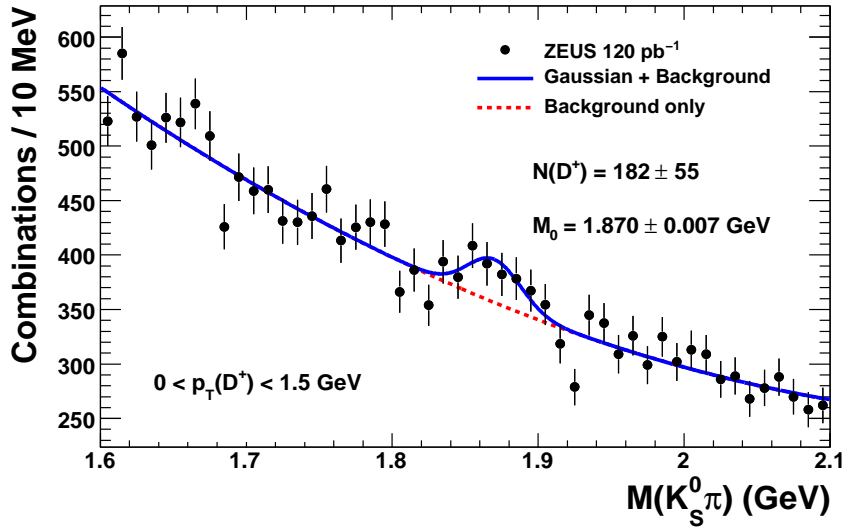,width=12cm}}\put(-210,202){\Huge\bfseries ZEUS}}
\caption{The $M(K^{0}_{S}\pi^{+})$ distribution (dots) for $D^{+}$ candidates in the region $0 < p_{T}(D^{+}) < 1.5\gev$. The reflection caused by the decay $D_{s}^{+}\to K^{0}_{S}K^{+}$ has been subtracted as described in the text. The solid curve represents a fit to the sum of a Gaussian signal and a background function, while the background contribution alone is given by the dashed curve.}
\label{fig:peak_dpm_onlyfirstbin}
\end{figure}

The signal in the region $0 < p_{T}(D^{+}) < 1.5\gev$ that was not accessible in previous measurements is shown in Fig.~\ref{fig:peak_dpm_onlyfirstbin}.

\boldmath
\subsection{Reconstruction of the decay $\Lambda_{c}^{+}\to pK^{0}_{S}$}
\unboldmath

The $\Lambda_{c}^{+}$ baryons were reconstructed from the decay channel $\Lambda_{c}^{+}\to pK^{0}_{S}$. In each event, $\Lambda_{c}^{+}$ candidates were formed from combinations of $K^{0}_{S}$ candidates reconstructed as described in Section~\ref{sec-strange-reco} with proton candidates. The proton-candidate selection used the energy-loss measurement in the CTD. Tracks fitted to the primary vertex with more than 40 hits were considered. The proton band was parametrised separately for positive and negative tracks from an examination of $dE/dx$ as a function of the momentum~\cite{thesis:roloff:2007}. The proton selection was checked by studying proton-candidate tracks from $\Lambda$ decays. To remove the region where the proton band completely overlaps the pion band, the proton momentum was required to be less than $1.5\gev$ and a cut on $dE/dx > 1.2$~mips was applied. Due to the proton selection described above, reflections from $D^{+}\to K^{0}_{S}\pi^{+}$ and $D_{s}^{+}\to K^{0}_{S}K^{+}$ decays are suppressed.

As a result of the cut on the proton momentum, there is no acceptance for $\Lambda_{c}^{+}$ baryons at very high $p_{T}(\Lambda_{c}^{+})$. Hence the measurement of the cross section for this decay channel was restricted to the region $0 < p_{T}(\Lambda_{c}^{+}) < 6\gev$.

\begin{figure}[hbtp]
\epsfysize=14cm
\centerline{\mbox{\epsfig{file=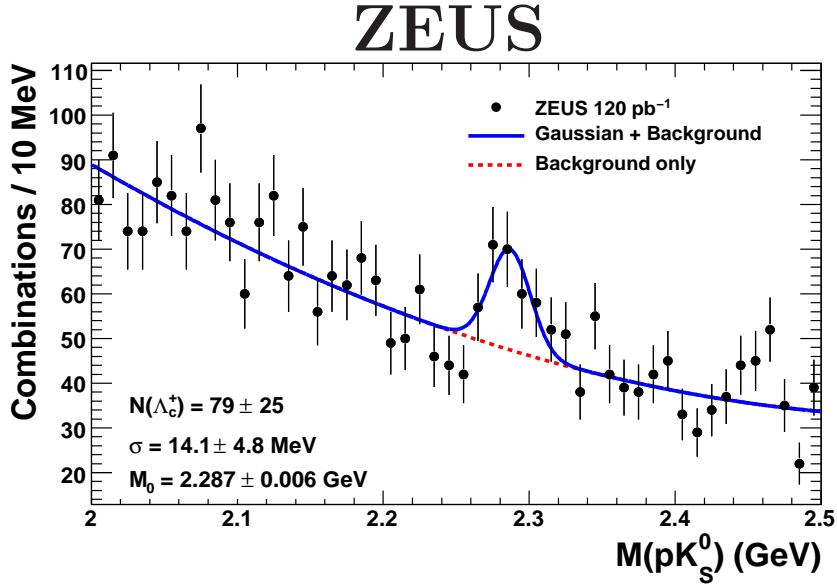,width=12cm}}\put(-210,202){\Huge\bfseries ZEUS}}
\caption{The $M(pK^{0}_{S})$ distribution (dots) for $\Lambda_{c}^{+}$ candidates in the region $0 < p_{T}(\Lambda_{c}^{+}) < 6\gev$. The solid curve represents a fit to the sum of a Gaussian signal and a background function, while the background contribution alone is given by the dashed curve.}
\label{fig:peak_k0p}
\end{figure}

Figure~\ref{fig:peak_k0p} shows the $M(pK^{0}_{S})$ distribution for the $\Lambda_{c}^{+}$ candidates. A clear signal is seen at the nominal value of the $\Lambda_{c}^{+}$ mass~\cite{Amsler:2008zzb}. The mass distribution was fitted to the sum of a Gaussian function describing the signal and the function defined in Eq.~(\ref{background-cheby}) to describe the non-resonant background. The number of reconstructed $\Lambda_{c}^{+}$ baryons yielded by the fit was $N(\Lambda_{c}^{+}) = 79 \pm 25$.

\boldmath
\subsection{Reconstruction of the decay $\Lambda_{c}^{+}\to \Lambda\pi^{+}$}
\unboldmath

The $\Lambda_{c}^{+}$ baryons were also reconstructed from the decay channel $\Lambda_{c}^{+}\to \Lambda\pi^{+}$. In each event, $\Lambda_{c}^{+}$ candidates were formed from combinations of $\Lambda$ candidates as described in Section~\ref{sec-strange-reco}, with further tracks assumed to be pions. The pion candidates were required to have $p_{T}(\pi^{+})/E_{T}^{\theta>10^{\circ}} > 0.05$. Only pion candidates with $dE/dx < 1.5$~mips were considered. To suppress combinatorial background further, the cut $\cos{\theta^{*}(\pi^{+})} < 0.8$ was imposed, where $\theta^{*}(\pi^{+})$ is the angle between the pion in the $\Lambda_{c}^{+}$ rest frame and the $\Lambda_{c}^{+}$ flight direction.

\begin{figure}[hbtp]
\epsfysize=14cm
\centerline{\mbox{\epsfig{file=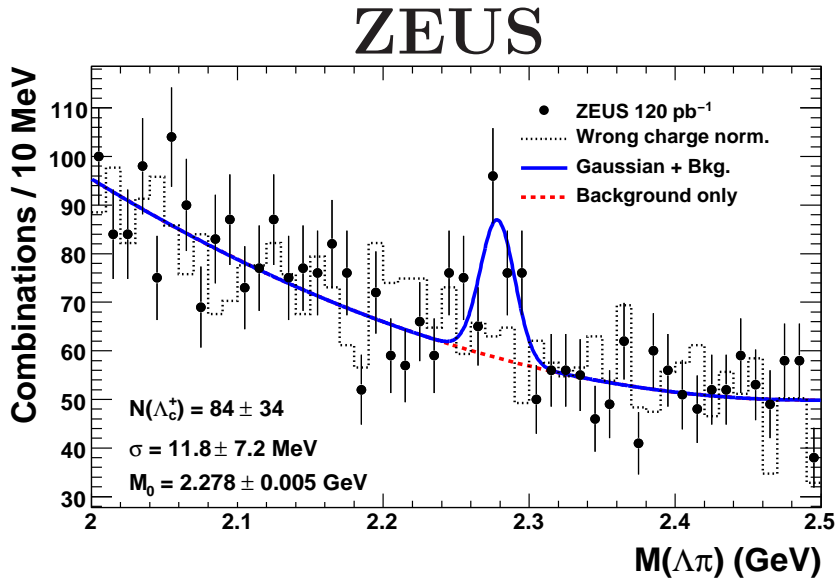,width=12cm}}\put(-210,202){\Huge\bfseries ZEUS}}
\caption{The $M(\Lambda\pi^{+})$ distribution (dots) for $\Lambda_{c}^{+}$ candidates. The solid curve represents a fit to the sum of a Gaussian signal and a background function, while the background contribution alone is given by the dashed curve. The dotted histogram shows the distribution of wrong-charge combinations (see text).}
\label{fig:peak_lambdapi}
\end{figure}

Figure~\ref{fig:peak_lambdapi} shows the $M(\Lambda\pi)$ distribution for the $\Lambda_{c}^{+}$ candidates. Wrong-charge combinations in the data sample, normalised to the right-charge combinations in the region outside the peak, are also shown. For wrong-charge combinations, the sum of the charges of the proton from the $\Lambda$ candidate and the further track is equal to zero. The data were fitted to the sum of a Gaussian function describing the signal and the background function defined in Eq.~(\ref{background-cheby}). The number of reconstructed $\Lambda_{c}^{+}$ baryons obtained from the fit was $N(\Lambda_{c}^{+}) = 84 \pm 34$.

\begin{figure}[hbtp]
\epsfysize=14cm
\centerline{\mbox{\epsfig{file=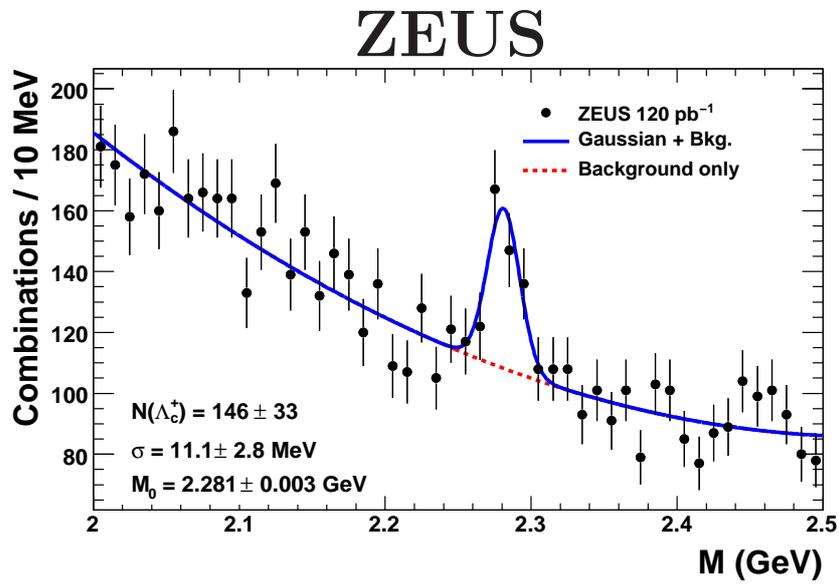,width=12cm}}\put(-210,202){\Huge\bfseries ZEUS}}
\caption{The invariant mass distribution (dots) for $\Lambda_{c}^{+}\to pK^{0}_{S}$ and $\Lambda_{c}^{+}\to \Lambda\pi^{+}$ candidates. The solid curve represents a fit to the sum of a Gaussian signal and a background function, while the background contribution alone is given by the dashed curve.}
\label{fig:peak_lambdac_combined}
\end{figure}

The signal-to-background ratio for both studied $\Lambda_{c}^{+}$ decay channels is similar. Figure~\ref{fig:peak_lambdac_combined} shows the invariant-mass spectrum containing both $\Lambda_{c}^{+}\to pK^{0}_{S}$ and $\Lambda_{c}^{+}\to \Lambda\pi^{+}$ candidates. The fit yielded $N(\Lambda_{c}^{+}) = 146 \pm 33$ candidates. This combined peak was not used to extract any cross sections or fragmentation fractions.

\section{Cross sections and acceptance corrections}
\label{sec-accep}

For a given observable, $Y$, the differential cross section in a bin $i$ was determined using
\begin{equation}
\frac {d\sigma_{i}}{dY} =
\frac {N_{i}(D^{+}) } {\mathcal {A}_{i} \cdot \mathcal {L} \cdot \mathcal {B} \cdot
\Delta Y_{i}},
\nonumber
\end{equation}
where $N_{i}(D^{+})$ is the number of reconstructed $D^{+}$ mesons in bin $i$ having size $\Delta Y_{i}$. The reconstruction acceptance, $\mathcal {A}_{i}$, takes into account migrations, efficiencies and QED radiative effects for the $i^{th}$ bin, $\mathcal {L}$ is the integrated luminosity and $\mathcal {B}$ is the branching ratio~\cite{Amsler:2008zzb} for the decay channel used in the reconstruction (see Table~\ref{tab-branching-ratios}). The total visible production cross sections were determined using
\begin{equation}
\sigma =
\frac {N(D^{+},\Lambda_{c}^{+}) } {\mathcal {A} \cdot \mathcal {L} \cdot \mathcal {B}},
\nonumber
\end{equation}
where $N(D^{+},\Lambda_{c}^{+})$ and $\mathcal {A}$ were determined for the whole kinematic range of the measurement. All acceptances were obtained from the Monte Carlo.

The $b$-quark contribution, predicted by the MC simulation, was subtracted from all measured cross sections. The {\sc Rapgap} prediction for beauty production was multiplied by two, in agreement with a previous ZEUS measurement of beauty production in DIS~\cite{pl:b599:173}. The subtraction of the $b$-quark contribution reduced the measured cross sections by $2-3\%$ for the $D^{+}$ and about $1\%$ for the $\Lambda_{c}^{+}$.

There is no sizeable acceptance for charmed hadrons in the transverse-momentum range $0 < p_{T}(D^{+}, \Lambda_{c}^{+}) < 0.5\gev$. Hence an extrapolation using the reference Monte Carlo was performed when the cross sections were extracted. For example, the extrapolation accounts for $6\%$ of the $D^{+}$ production in the full kinematic range of the measurement and for $11\%$ of the $D^{+}$ production in the restricted range $0 < p_{T}(D^{+}) < 1.5\gev$.

\section{Systematic uncertainties}
\label{sec-syst}

The systematic uncertainties of the measured cross sections and fragmentation fractions were determined by changing the analysis procedure and repeating all calculations. In the measurement of the differential and total cross sections, the following groups of systematic uncertainty sources were considered. The effects on the total cross sections are shown in parentheses ($D^{+}$; $\Lambda_{c}^{+}\to pK^{0}_{S}$; $\Lambda_{c}^{+}\to \Lambda\pi^{+}$):

\begin{itemize}

\item[$\bullet$] {\{$\delta_{1}$\} event and DIS selection ($^{+4\%}_{-3\%}$; $^{+1\%}_{-2\%}$; $^{+8\%}_{-4\%}$). The following cut variations were applied to data and MC simultaneously:}

\begin{itemize}

\item the cut on $y_{{\rm JB}}$ was changed to $y_{{\rm JB}} > 0.03$;

\item the cut on the scattered electron energy $E_{e}^{'}$ was changed to $E_{e}^{'} > 11\gev$;

\item the cuts on $\delta$ were changed by $+2\gev$;

\item the cut on $|Z_{\textnormal{vertex}}|$ was changed to $|Z_{\textnormal{vertex}}| < 45\cm$;

\item additionally, a box cut of $26 \times 14\cm^{2}$ was used for all electron candidates without an SRTD requirement;

\end{itemize}

\item[$\bullet$] {\{$\delta_{2}$\} $Q^{2}$ and $x$ reconstruction ($<\!\!1\%$; $-3\%$; $-6\%$). The DA method was used for the reconstruction of $Q^{2}$ and $x$ instead of the electron method;}

\item[$\bullet$] {\{$\delta_{3}$\} energy scale ($\pm 2\%$; $^{+3\%}_{-4\%}$; $^{+2\%}_{-4\%}$). To account for the uncertainty of the absolute CAL energy scale, the energy of the scattered electron was raised and lowered by $1\%$ and $E_{T}^{\theta>10^{\circ}}$ was raised and lowered by $2\%$. These variations were only applied to the MC;}

\item[$\bullet$] {\{$\delta_{4}$\} model dependence of the acceptance corrections:}

\begin{itemize}

\item the process $cg\to cg$ was included in the {\sc Rapgap} MC sample ($+5\%$; $+3\%$; $+9\%$);

\item the MC samples were not reweighted in $p_{T}(D^{+}, D_{s}^{+}, \Lambda_{c}^{+})$ ($-17\%$; $-6\%$; $-21\%$);

\end{itemize}

\item[$\bullet$] {\{$\delta_{5}$\} uncertainty of the beauty subtraction ($^{+1\%}_{-3\%}$; $\pm1\%$; $<\!\! 1\%$). This was determined by varying the subtracted $b$-quark contributions by a factor 2;}

\item[$\bullet$] {\{$\delta_{6}$\} uncertainty of the signal extraction procedure ($^{+12\%}_{-9\%}$; $^{+14\%}_{-5\%}$; $^{+24\%}_{-8\%}$):}

\begin{itemize}

\item the fit was repeated changing the invariant mass window of $1.6 - 2.1\gev$ by $\pm 50\mev$ on both sides for $D^{+}\to K^{0}_{S}\pi^{+}$ decays. Similarly, the considered invariant mass region of $2.0 - 2.5\gev$ was changed by $\pm 50\mev$ for $\Lambda_{c}^{+}\to pK^{0}_{S}$ decays and by $\pm 30\mev$ for the channel $\Lambda_{c}^{+}\to \Lambda\pi^{+}$;

\item the choice of the background function was assigned an uncertainty of $\pm 5\%$. This value was estimated by comparing the fit results obtained using different choices for the background function, such as polynominals of different orders or exponential functions;

\item for differential cross sections, the assumed Gaussian width ratios were varied by $\pm 10\%$;

\end{itemize}

\item[$\bullet$] {\{$\delta_{7}$\} uncertainty in the luminosity measurement of $\pm 2.0\%$.}

\end{itemize}

The following uncertainty was considered only for the decays $D^{+}\to K^{0}_{S}\pi^{+}$ and $\Lambda_{c}^{+}\to K^{0}_{S}p$:

\begin{itemize}

\item[$\bullet$] {\{$\delta_{8}$\} $K^{0}_{S}$ reconstruction ($+2\%$; $+1\%$; $-$). Since the MC signal had a narrower width than observed in the data, the invariant-mass window for the $K^{0}_{S}$ candidate selection was reduced to $0.486 < M(\pi^{+}\pi^{-}) < 0.510\gev$ in the MC only.}

\end{itemize}

The following source of uncertainty was considered only for the decay $D^{+}\to K^{0}_{S}\pi^{+}$:

\begin{itemize}

\item[$\bullet$] {\{$\delta_{9}$\} uncertainty of the reflection subtraction($\pm5\%$; $-$; $-$). The normalisation of the $D_{s}^{+}$ reflection was changed by the uncertainty of $R$ (see Eq.~(\ref{ratio_reflection})) due to the uncertainties of the fragmentation fractions and branching ratios used in the calculation.}

\end{itemize}

The following source of uncertainty was considered only for the decay $\Lambda_{c}^{+}\to K^{0}_{S}p$:

\begin{itemize}

\item[$\bullet$] {\{$\delta_{10}$\} proton reconstruction ($-$; $-14\%$; $-$). The following checks were performed:}

\begin{itemize}

\item{the number of hits required for the proton candidates was lowered to 32;}

\item the uncertainty of the $dE/dx$ simulation for low-momentum protons was evaluated changing the parametrisation of the proton band~\cite{thesis:roloff:2007};

\item the cut on the energy loss was lowered to $dE/dx > 1.15$~mips.

\end{itemize}

\end{itemize}

The following source of uncertainty was considered only for the decay $\Lambda_{c}^{+}\to \Lambda\pi^{+}$:

\begin{itemize}

\item[$\bullet$] {\{$\delta_{11}$\} $\Lambda$ reconstruction ($-$; $-$; $+4\%$). Since the MC signals had a narrower width than observed in the data, the invariant-mass window for the $\Lambda$ candidate selection was reduced to $1.113 < M(p\pi) < 1.120\gev$ in the MC only.}

\end{itemize}

Contributions from the different systematic uncertainties were calculated and added in quadrature separately for positive and negative variations. These estimates were made in each bin in which the differential cross sections were measured. Uncertainties due to those on the luminosity measurement and branching ratios were only included in the measured $D^{+}$ and $\Lambda_{c}^{+}$ total cross sections. For differential cross sections, these uncertainties are not included.

As an additional check, the $dE/dx$ efficiency for pions and protons was verified directly in the data using $K^{0}_{S}$ and $\Lambda$ decays. For the $D^{+}\to K^{0}_{S}\pi^{+}$ decay channel, the effect of the $dE/dx$ cut on the pion candidate tracks was very small and the result changed only marginally when the cut was released.

The average cross sections obtained from the two different running periods ($\sqrt{s}=300$ and $318\gev$) are expressed in terms of cross sections at $\sqrt{s}=318\gev$. This involves a typical correction of +1\% determined using HVQDIS.

\section{Results}
\label{sec-results}

Charm hadron cross sections were measured using the reconstructed $D^{+}$ and $\Lambda^{+}_{c}$ signals (see Section~\ref{sec-charm-reco}) in the kinematic range $0 < p_{T}(D^{+}, \Lambda_{c}^{+}) < 10\gev$, $|\eta(D^{+},\Lambda_{c}^{+})| < 1.6$, $1.5 < Q^{2} < 1000\gev^{2}$ and $0.02 < y < 0.7$.

In addition to the statistical and systematic uncertainties, a third set of uncertainties is quoted for the measured cross sections and charm fragmentation fractions, due to the propagation of the relevant branching-ratio uncertainties (Table~\ref{tab-branching-ratios}).

\boldmath
\subsection{$D^{+}$ cross sections}
\unboldmath

The following total visible cross section for $D^{+}$ mesons was measured:
\begin{equation}
\sigma(D^{+}) = 25.7 \pm 4.1 ~(\rm stat.) ~^{+3.8}_{-5.2} ~(\rm syst.) \pm 0.8 ~(\rm br.) \rm ~nb.
\nonumber
\end{equation}
The corresponding prediction from HVQDIS is $\sigma(D^{+}) = 12.7 ~^{+3.8}_{-4.1} \rm ~nb$. The measured and predicted cross sections are in agreement to better than two standard deviations.

To allow a direct comparison to a recent measurement of $D^{+}$ production by the ZEUS collaboration using a lifetime tag~\cite{Chekanov:2008yd}, the cross section was extracted for the kinematic region defined by $1.5 < p_{T}(D^{+}) < 15\gev$, $|\eta(D^{+})| < 1.6$, $5.0 < Q^{2} < 1000\gev^{2}$ and $0.02 < y < 0.7$. The measurements using different decay channels and different techniques were found to be consistent.

\begin{figure}[hbtp]
\begin{center}
\includegraphics[width=0.49\textwidth]{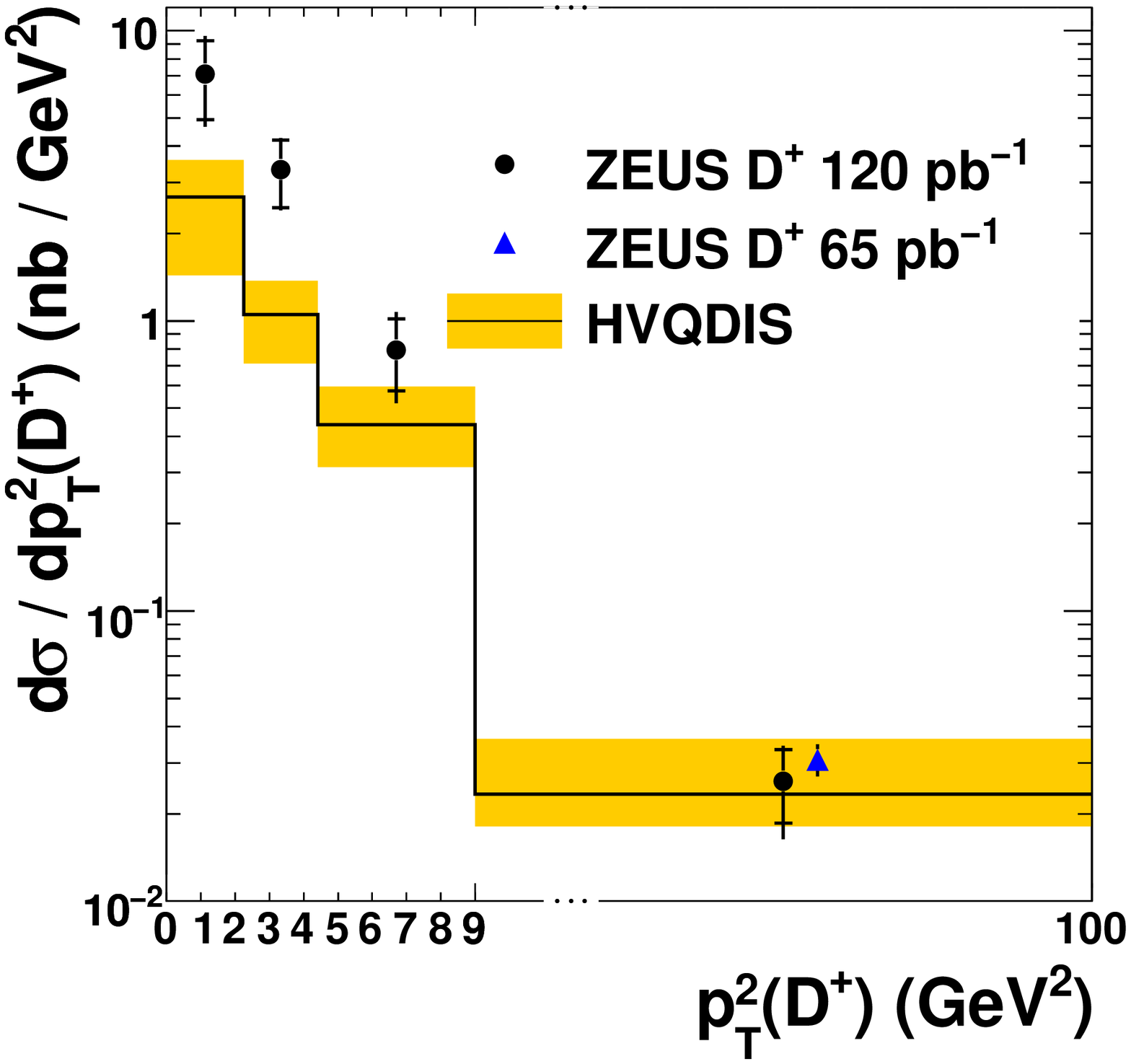}
\put(-48,188){\makebox(0,0)[tl]{\large (a)}}
\includegraphics[width=0.49\textwidth]{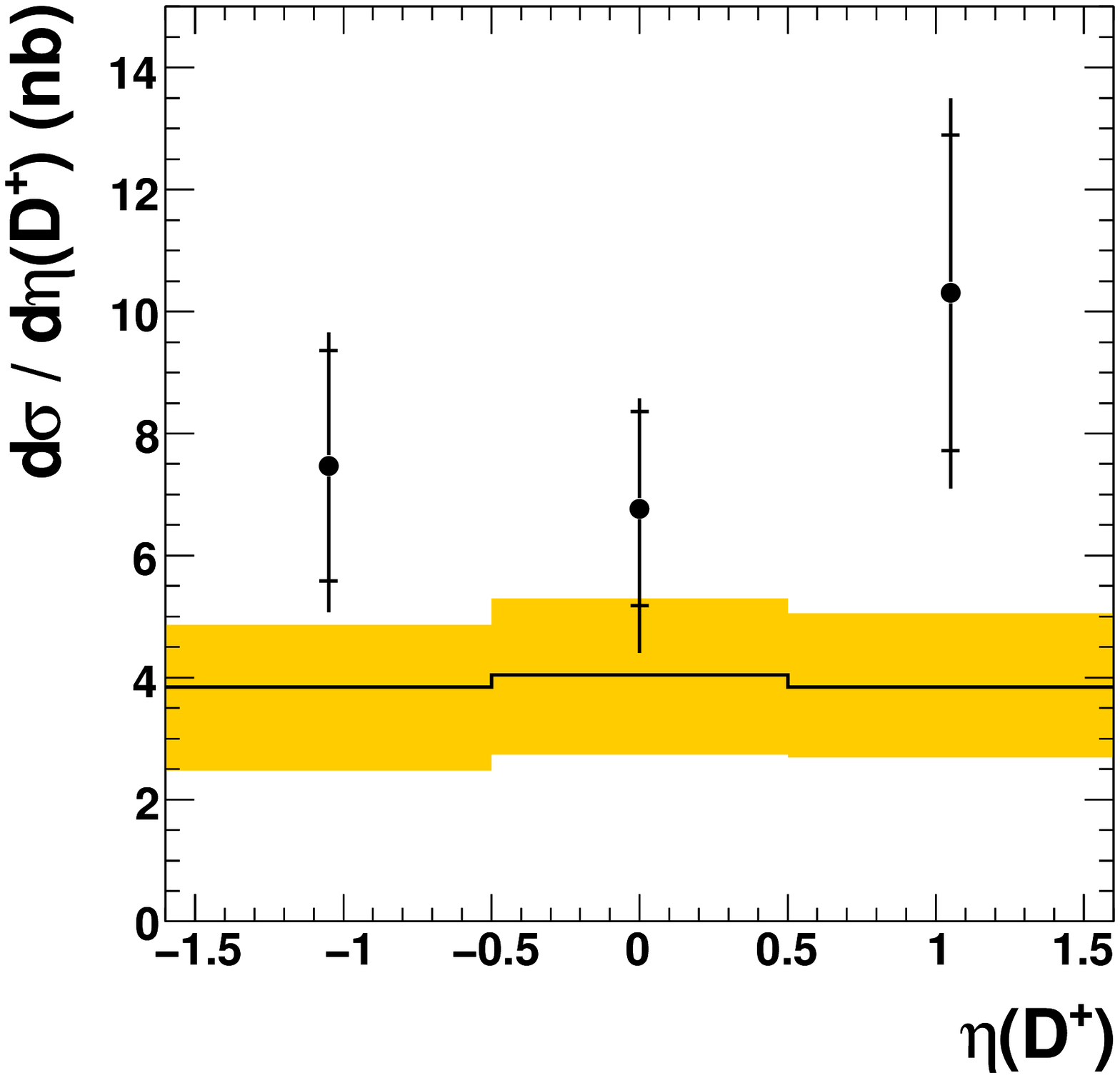}
\put(-48,188){\makebox(0,0)[tl]{\large (b)}}
\put(-260,202){\Huge\bfseries ZEUS}

\includegraphics[width=0.49\textwidth]{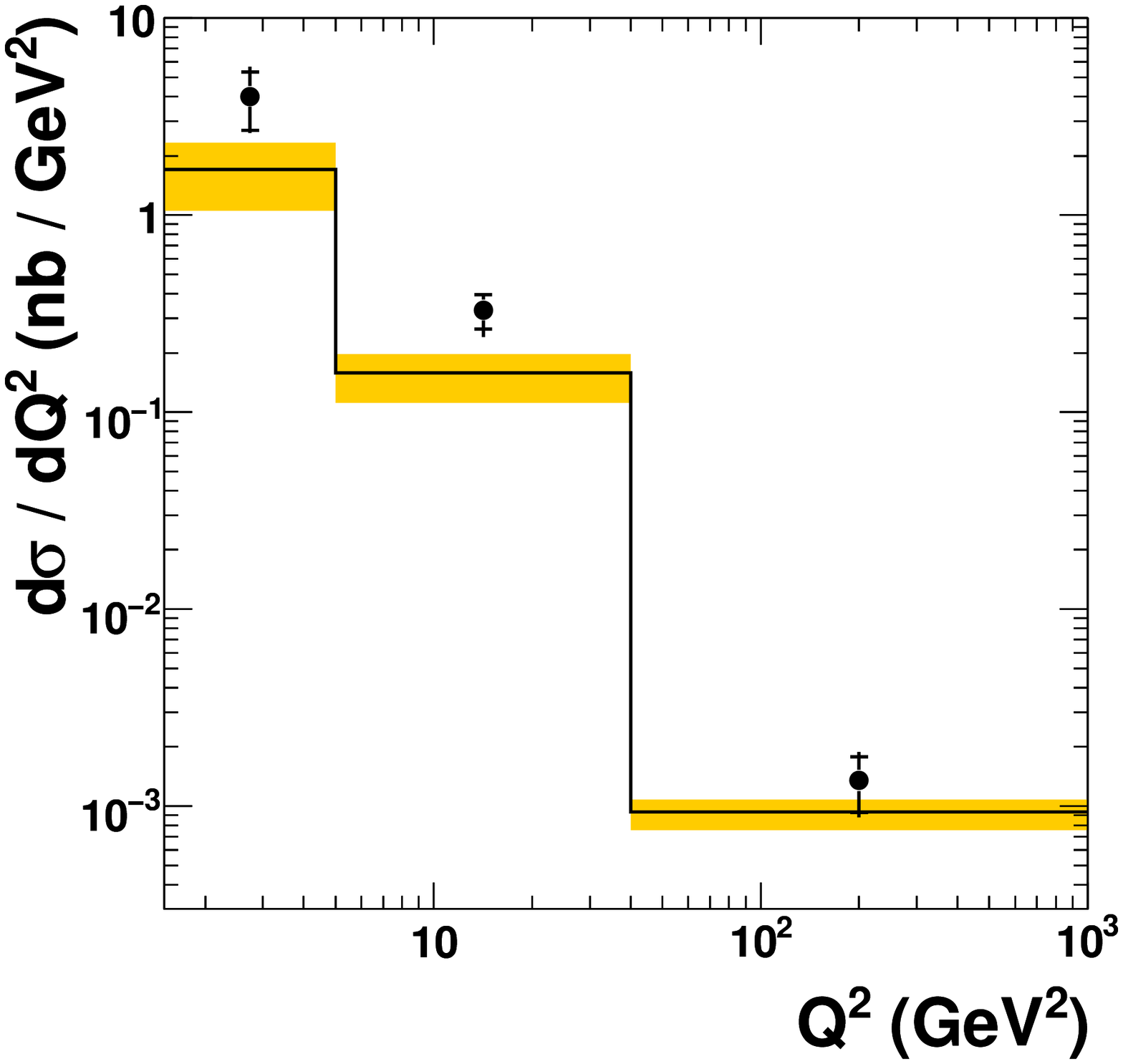}
\put(-48,188){\makebox(0,0)[tl]{\large (c)}}
\includegraphics[width=0.49\textwidth]{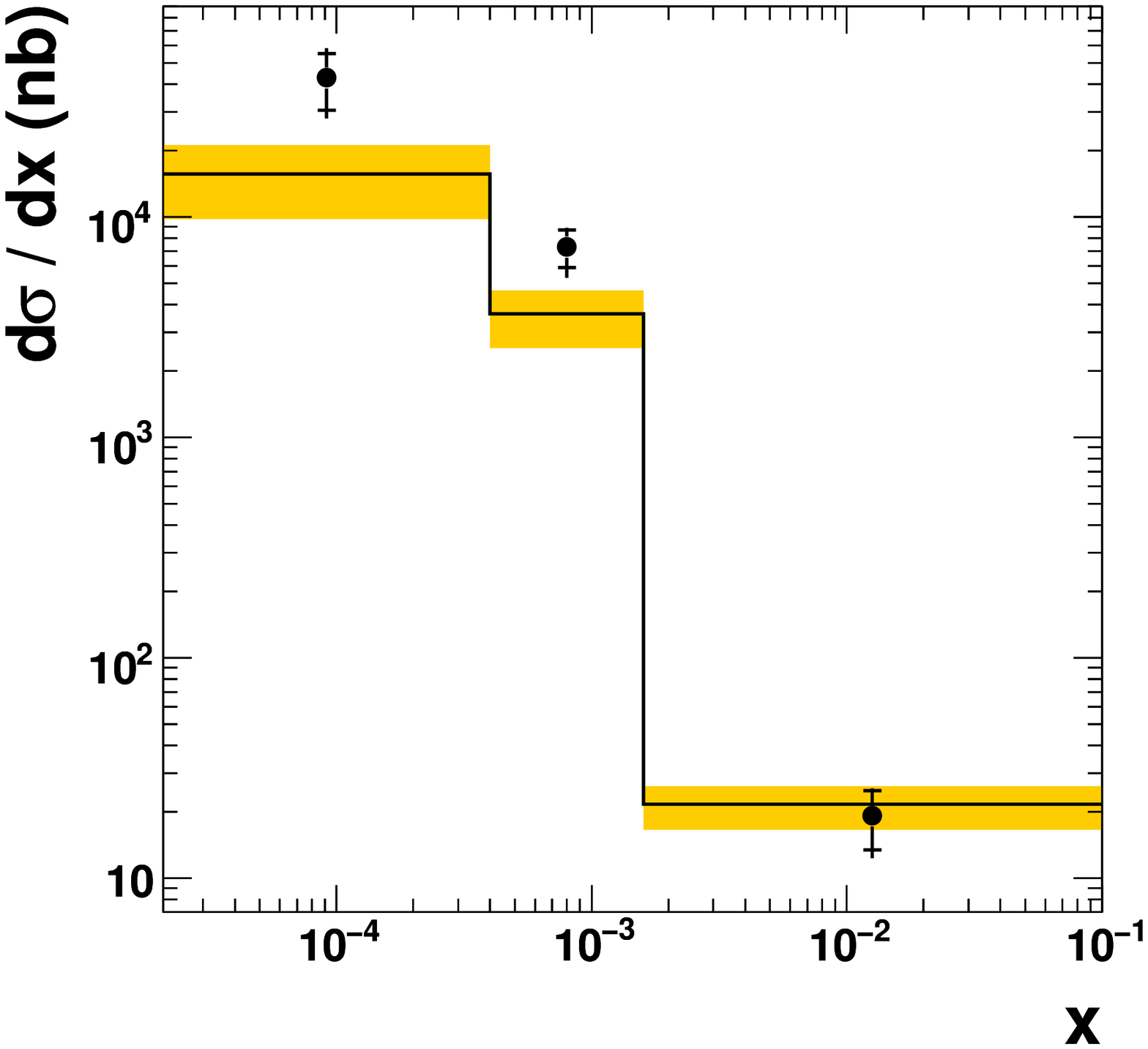}
\put(-48,188){\makebox(0,0)[tl]{\large (d)}}
\caption{Differential $D^{+}$ cross sections as a function of (a) $p_{T}^{2}(D^{+})$, (b) $\eta(D^{+})$, (c) $Q^{2}$ and (d) $x$ compared to the NLO QCD calculation of HVQDIS. The measured cross sections are shown as dots and the triangle represents a previous ZEUS result. The $X$-axis in (a) is broken. The inner error bars show the statistical uncertainties and the outer error bars show the statistical and systematic uncertainties added in quadrature. The band shows the estimated theoretical uncertainty of the HVQDIS calculation.}
\label{fig:xsections_dpm}
\end{center}
\end{figure}

The differential cross sections as functions of $p_{T}^{2}(D^{+})$, $\eta(D^{+})$, $x$ and $Q^{2}$ are shown in Fig.~\ref{fig:xsections_dpm} and given in Table~\ref{tab:dplus_xsections_pt2_eta_q2_x}. The cross sections in $Q^{2}$ and $x$ fall by about three orders of magnitude, while the cross section in $p_{T}^{2}(D^{+})$ falls by about two orders of magnitude in the measured region. There is no significant dependence of the cross section on $\eta(D^{+})$. The HVQDIS predictions describe the shape of all measured differential cross sections reasonably well. The differential cross section in $p_{T}^{2}(D^{+})$ is compared to a previous ZEUS result~\cite{Chekanov:2007ch} for $p_{T}^{2}(D^{+}) > 9\gev^{2}$. The two measurements are in good agreement.

\boldmath
\subsection{$\Lambda_{c}^{+}$ cross sections and fragmentation fractions}
\unboldmath

The following $\Lambda_{c}^{+}$ cross sections were measured:

\begin{itemize}

\item{using the decay channel $\Lambda_{c}^{+}\to pK^{0}_{S}$ in the restricted range $0 < p_{T}(\Lambda_{c}^{+}) < 6\gev$:}
\begin{equation}
\sigma(\Lambda_{c}^{+}) = 14.9 \pm 4.9 ~(\rm stat.) ~^{+2.2}_{-2.6} ~(\rm syst.) \pm 3.9 ~(\rm br.) \rm ~nb;
\nonumber
\end{equation}
\item{using the decay channel $\Lambda_{c}^{+}\to \Lambda\pi^{+}$:}
\begin{equation}
\sigma(\Lambda_{c}^{+}) = 14.0 \pm 5.8 ~(\rm stat.) ~^{+3.8}_{-3.3} ~(\rm syst.) \pm 3.7 ~(\rm br.) \rm ~nb.
\nonumber
\end{equation}
\end{itemize}

To compare and combine both measurements, the value obtained for the decay channel $\Lambda_{c}^{+}\to pK^{0}_{S}$ was multiplied by $1.01 \pm 0.01$ to extrapolate to the full kinematic region considered in this paper. The cross sections obtained using different decay channels are in good agreement. To extract the $\Lambda_{c}^{+}$ fragmentation fraction, the measurements were combined taking into account all systematic uncertainties and their correlations:
\begin{equation}
\sigma_{\rm combined}(\Lambda_{c}^{+}) = 14.7 \pm 3.8 ~(\rm stat.) ~^{+2.1}_{-2.2} ~(\rm syst.) \pm 3.9 ~(\rm br.) \rm ~nb.
\nonumber
\end{equation}
The uncertainty of the branching ratio was treated as partially correlated since both branching ratios, $\mathcal{B}(\Lambda_{c}^{+}\to pK^{0}_{S})$ and $\mathcal{B}(\Lambda_{c}^{+}\to \Lambda\pi^{+})$, were measured relative to the decay mode $\Lambda_{c}^{+}\to pK^{-}\pi^{+}$~\cite{Amsler:2008zzb}.

The fragmentation fraction $f(c\to \Lambda_{c}^{+})$ can be calculated using the $D^{+}$ cross section:
\begin{equation}
f(c\to \Lambda_{c}^{+}) = \frac{\sigma(\Lambda_{c}^{+})}{\sigma(D^{+})} \cdot f(c\to D^{+}).
\label{f_lambdac}
\end{equation}
In a previous ZEUS publication~\cite{Chekanov:2007ch} $f(c\to D^{+})$ was defined as:
\begin{equation}
f(c\to D^{+}) = \frac{\sigma^{0}(D^{+})}{\sigma^{0}(D^{+}) + \sigma^{0}(D^{0}) + \sigma^{0}(D_{s}^{+})} \cdot \left[ 1 - 1.14 \cdot f(c\to \Lambda_{c}^{+}) \right],
\label{f_dplus}
\end{equation}
where $\sigma^{0}(D^{+})$, $\sigma^{0}(D^{0})$ and $\sigma^{0}(D_{s}^{+})$ are the cross sections for $p_{T}(D) > 3\gev$. The factor $1.14$ takes into account the production of charm-strange baryons~\cite{Chekanov:2007ch}. For $D^{+}$ and $D^{0}$ mesons the equivalent cross sections (as described elsewhere~\cite{Chekanov:2005mm}) were used. Combining Eqs.~(\ref{f_lambdac}) and~(\ref{f_dplus}) yields:
\begin{equation}
f(c\to \Lambda_{c}^{+}) = \frac{\sigma(\Lambda_{c}^{+}) \cdot \sigma^{0}(D^{+})}{\sigma(D^{+}) \cdot (\sigma^{0}(D^{+}) + \sigma^{0}(D^{0}) + \sigma^{0}(D_{s}^{+})) + 1.14 \; \sigma(\Lambda_{c}^{+}) \cdot \sigma^{0}(D^{+})}
\nonumber
\end{equation}
Since the cross sections $\sigma(D^{+})$ and $\sigma(\Lambda_{c}^{+})$ were measured down to $p_{T}(D^{+},\Lambda_{c}^{+}) = 0\gev$, no treatment of the different transverse momentum distributions for $D^{+}$ and $\Lambda_{c}^{+}$ hadrons was necessary. The measured value:
\begin{equation}
f(c\to \Lambda_{c}^{+}) = 0.117 \pm 0.033 ~(\rm stat.) ~^{+0.026}_{-0.022} ~(\rm syst.) \pm 0.027 ~(\rm br.),
\nonumber
\end{equation}
is compared to previous measurements in Table~\ref{tab:fragmentation_fraction}. The result is consistent with a previous ZEUS measurement in the photoproduction regime~\cite{Chekanov:2005mm} and with the $e^{+}e^{-}$ average value.

\section{Conclusions}
\label{sec-concl}

Open-charm production in $ep$ collisions at HERA has been measured in deep inelastic scattering using three decay channels. The presence of a neutral strange hadron in the final state allowed the measurement to be extended to very low transverse momenta of the reconstructed charmed hadrons. The total visible and differential cross sections for $D^{+}$ production are in reasonable agreement with NLO QCD predictions. The measured $D^{+}$ cross sections are consistent with previous ZEUS results. The fragmentation fraction $f(c\to \Lambda_{c}^{+})$ has been measured for the first time at HERA in deep inelastic scattering. The result obtained from a combination of two decay channels is consistent with a previous measurement performed in the photoproduction regime and with the average $e^{+}e^{-}$ value.

\section{Acknowledgements}
\label{sec-acknow}

We appreciate the contributions to the construction and maintenance of the ZEUS detector of many people who are not listed as authors. The HERA machine group and the DESY computing staff are especially acknowledged for their success in providing excellent operation of the collider and the data-analysis environment. We thank the DESY directorate for their strong support and encouragement.

\vfill\eject

{
\def\bibname{\Large\bf References}
\def\refname{\Large\bf References}
\pagestyle{plain}
\ifzeusbst
  \bibliographystyle{./BiBTeX/bst/l4z_default}
\fi
\ifzdrftbst
  \bibliographystyle{./BiBTeX/bst/l4z_draft}
\fi
\ifzbstepj
  \bibliographystyle{./BiBTeX/bst/l4z_epj}
\fi
\ifzbstnp
  \bibliographystyle{./BiBTeX/bst/l4z_np}
\fi
\ifzbstpl
  \bibliographystyle{./BiBTeX/bst/l4z_pl}
\fi
{\raggedright
\bibliography{./BiBTeX/user/syn.bib,%
             ./BiBTeX/bib/l4z_articles.bib,%
             ./BiBTeX/bib/l4z_books.bib,%
             ./BiBTeX/bib/l4z_conferences.bib,%
             ./BiBTeX/bib/l4z_h1.bib,%
             ./BiBTeX/bib/l4z_misc.bib,%
             ./BiBTeX/bib/l4z_old.bib,%
             ./BiBTeX/bib/l4z_preprints.bib,%
             ./BiBTeX/bib/l4z_replaced.bib,%
             ./BiBTeX/bib/l4z_temporary.bib,%
             ./BiBTeX/bib/l4z_zeus.bib}}
}
\vfill\eject

\begin{table}
\begin{center}
\begin{tabular}{|c|c|}
\hline
Decay mode & Branching ratio [\%] \\ 
\hline\hline
$D^{+}\to K^{0}_{S}\pi^{+}\to \pi^{+}\pi^{-}\pi^{+}$ & $1.00 \pm 0.03$ \\
$D^{+}_{s}\to K^{+}K^{0}_{S}\to K^{+}\pi^{+}\pi^{-}$ & $1.03 \pm 0.06$ \\
$\Lambda^{+}_{c}\to pK^{0}_{S}\to p\pi^{+}\pi^{-}$ & $0.80 \pm 0.21$ \\
$\Lambda^{+}_{c}\to \Lambda\pi^{+}\to p\pi^{-}\pi^{+}$ & $0.68 \pm 0.18$ \\
\hline
\end{tabular}
\caption{Branching ratios of the charmed hadron decays~\protect\cite{Amsler:2008zzb}.}
\label{tab-branching-ratios}
\end{center}
\end{table}

\begin{table}
\begin{center}
\begin{tabular}[width = \textwidth]{|r@{, }l|r@{.}l@{    }|r@{.}l@{    }|r@{.}l@{    }r@{.}l|}
\hline
\multicolumn{2}{|c|}{$p_{T}^{2}(D^{+})$ bin} & \multicolumn{2}{c|}{$d\sigma/dp_{T}^{2}(D^{+})$} & \multicolumn{2}{c|}{$\Delta_{\rm stat}$}& \multicolumn{4}{c|}{$\Delta_{\rm syst}$} \\
\multicolumn{2}{|c|}{($\gev^{2}$)} & \multicolumn{2}{c|}{(nb/$\gev^{2}$)} & \multicolumn{2}{c|}{(nb/$\gev^{2}$)} & \multicolumn{4}{c|}{(nb/$\gev^{2}$)} \\
\hline
\hline
0 & 2.25 & 7&1 & $\pm$2&1 & +1&3 & $-$1&1 \\
2.25 & 4.41 & 3&3 & $\pm$0&9 & +0&4 & $-$0&3 \\
4.41 & 9.0 & 0&80 & $\pm$0&22 & +0&17 & $-$0&16 \\
9.0 & 100.0 & 0&026 & $\pm$0&007 & +0&004 & $-$0&006 \\
\hline
\hline
\multicolumn{2}{|c|}{$\eta(D^{+})$ bin} & \multicolumn{2}{c|}{$d\sigma/d\eta(D^{+})$} & \multicolumn{2}{c|}{$\Delta_{\rm stat}$}& \multicolumn{4}{c|}{$\Delta_{\rm syst}$}\\
\multicolumn{2}{|c|}{} & \multicolumn{2}{c|}{(nb)} & \multicolumn{2}{c|}{(nb)} & \multicolumn{4}{c|}{(nb)} \\
\hline
\hline
$-$1.6 & $-$0.5 & 7&5 & $\pm$1&9 & +1&1 & $-$1&5 \\
$-$0.5 & 0.5 & 6&8 & $\pm$1&6 & +0&9 & $-$1&8 \\
0.5 & 1.6 & 10&3 & $\pm$2&6 & +1&9 & $-$1&9 \\
\hline
\hline
\multicolumn{2}{|c|}{$Q^{2}$ bin} & \multicolumn{2}{c|}{$d\sigma/dQ^{2}$} & \multicolumn{2}{c|}{$\Delta_{\rm stat}$}& \multicolumn{4}{c|}{$\Delta_{\rm syst}$}\\
\multicolumn{2}{|c|}{($\gev^{2}$)} & \multicolumn{2}{c|}{(nb/$\gev^{2}$)} & \multicolumn{2}{c|}{(nb/$\gev^{2}$)} & \multicolumn{4}{c|}{(nb/$\gev^{2}$)} \\
\hline
\hline
1.5 & 5.0 & 4&0 & $\pm$1&3 & +1&0 & $-$0&5 \\
5.0 & 40.0 & 0&33 & $\pm$0&06 & +0&03 & $-$0&06 \\
40.0 & 1000.0 & 0&0013 & $\pm$0&0004 & +0&0003 & $-$0&0002 \\
\hline
\hline
\multicolumn{2}{|c|}{$x$ bin} & \multicolumn{2}{c|}{$d\sigma/dx$} & \multicolumn{2}{c|}{$\Delta_{\rm stat}$}& \multicolumn{4}{c|}{$\Delta_{\rm syst}$}\\
\multicolumn{2}{|c|}{} & \multicolumn{2}{c|}{(nb)} & \multicolumn{2}{c|}{(nb)} & \multicolumn{4}{c|}{(nb)} \\
\hline
\hline
0.000021 & 0.0004 & \multicolumn{2}{l|}{43000} & \multicolumn{2}{l|}{$\pm$12000} & \multicolumn{2}{l}{+9000} & \multicolumn{2}{l|}{$-$8000} \\
0.0004 & 0.0016 & \multicolumn{2}{l|}{7300} & \multicolumn{2}{l|}{$\pm$1400} & \multicolumn{2}{l}{+800} & \multicolumn{2}{l|}{$-$1400} \\
0.0016 & 0.1 & \multicolumn{2}{l|}{19.2} & \multicolumn{2}{l|}{$\pm$5.7} & \multicolumn{2}{l}{+2.8} & \multicolumn{2}{l|}{$-$3.7} \\
\hline
\end{tabular}
\caption{Measured $D^{+}$ cross sections as a function of $p_{T}^{2}(D^{+})$, $\eta(D^{+})$, $Q^{2}$ and $x$ for $1.5 < Q^{2} < 1000\gev^{2}$, $0.02 < y < 0.7$, $0 < p_{T}(D^{+}) < 10\gev$ and $|\eta(D^{+})| < 1.6$. The statistical and systematic uncertainties are shown separately. The cross sections have further uncertainties of $3\%$ from the $D^{+} \rightarrow K^{0}_{S}\pi^{+}\to \pi^{+}\pi^{-}\pi^{+}$ branching ratio, and $2\%$ from the uncertainty in the luminosity measurement.}
\label{tab:dplus_xsections_pt2_eta_q2_x}
\end{center}
\end{table}

\begin{table}
\begin{center}
\begin{tabular}{|c|c|}
\hline
& $f(c\to \Lambda_{c}^{+})$ \\
\hline
\hline
ZEUS (DIS) & $0.117 \pm 0.033 ~({\rm stat.}) ~^{+0.026}_{-0.022} ~({\rm syst.}) \pm 0.027 ~({\rm br.})$ \\
\hline
ZEUS ($\gamma p$)~\cite{Chekanov:2005mm} & $0.144 \pm 0.022 ~({\rm stat.}) ~^{+0.013}_{-0.022} ~({\rm syst.}) ~^{+0.037}_{-0.025} ~({\rm br.})$ \\
\hline
combined $e^{+}e^{-}$ data & $0.076 \pm 0.007 ~({\rm stat.\oplus syst.}) ~^{+0.027}_{-0.016} ~({\rm br.})$ \\
\hline
\end{tabular}
\caption{The fraction of $c$ quarks hadronising to a $\Lambda_{c}^{+}$ baryon, $f(c\to \Lambda_{c}^{+})$.}
\label{tab:fragmentation_fraction}
\end{center}
\end{table}

%
%
\end{document}